\documentclass[a4paper,fleqn,usenatbib]{mnras}

\usepackage{mathptmx}

\usepackage[T1]{fontenc}
\usepackage{ae,aecompl}

\usepackage{graphicx}	
\usepackage{amsmath}	
\usepackage{amssymb}	
\usepackage{fixltx2e}
\usepackage{hyperref}
\usepackage{threeparttable}
\usepackage{multirow}
\usepackage{booktabs}

\newcommand{\Lsol}{\,$L_{\odot}$\,}
\newcommand{\Msol}{\,$M_{\odot}$\,}
\newcommand{\Rsol}{\,$R_{\odot}$\,}
\newcommand{\yr}{\,yr$^{-1}$\,}
\newcommand{\kms}{\,km\,s$^{-1}$\,}

\title[The photometric variability of magnetic massive stars]{Modelling the photometric variability of magnetic massive stars with the Analytical Dynamical Magnetosphere model}

\author[M. S. Munoz et al.]{
	M. S. Munoz,$^{1}$\thanks{E-mail: 16msm5@queensu.ca}
	G. A. Wade,$^{2}$
	Y. Naz\'e,$^{3}$\thanks{FNRS research associate}
	J. Puls,$^{4}$
    S. Bagnulo$^{5}$
    M. K. Szyma\'nski$^{6}$    
	\\
	$^{1}$Department of Physics, Engineering Physics and Astronomy, Queen's University, 64 Bader lane, Kingston, K7L 3N6, ON Canada\\
	$^{2}$Department of Physics and Space Science, Royal Military College of Canada, PO Box 17000, Stn Forces, Kingston, K7K 7B4, \\ ON, Canada\\
	$^{3}$Groupe d'Astrophysique des Hautes \'Energies, Institut d'Astrophysique et de G\'eophysique Department AGO, Universit\'e de Li\`ege,\\ All\'ee du 6 Ao\^ut 19c
	B5c, 4000 Sart Tilman, Belgium  \\
	$^{4}$LMU M\"unchen, Universit\"atssternwarte
Scheinerstr. 1, D-81679 M\"unchen, Germany \\
	$^{5}$Armagh Observatory and Planetarium, College Hill, Armagh BT61 9DG, U.K.  \\
	$^{6}$Astronomical Observatory, University of Warsaw, Aleje Ujazdowskie 4, 00-478 Warszawa, Poland\\
}

\date{Accepted XXX. Received YYY; in original form ZZZ}

\pubyear{2018}

\begin{document}
	\label{firstpage}
	\pagerange{\pageref{firstpage}--\pageref{lastpage}}
	\maketitle
	
	\begin{abstract}
 In this paper, we investigate the photometric variability of magnetic O-type stars. Such stars possess oblique, predominantly dipolar magnetic fields that confine their winds roughly axisymmetrically about the magnetic equator, thus forming a magnetosphere. We interpret their photometric variability as phase-dependent magnetospheric occultations. For massive star winds dominated by electron scattering opacity in the optical and NIR, we can compute synthetic light curves from simply knowing the magnetosphere's mass density distribution. We exploit the newly-developed Analytical Dynamical Magnetosphere model (ADM) in order to obtain the predicted circumstellar density structures of magnetic O-type stars. The simplicity in our light curve synthesis model allows us to readily conduct a parameter space study. For validation purposes, we first apply our algorithm to HD 191612, the prototypical Of?p star. Next, we attempt to model the photometric variability of the Of?p-type stars identified in the Magellanic Clouds using OGLE photometry. We evaluate the compatibility of the ADM predictions with the observed photometric variations, and discuss the magnetic field properties that are implied by our modelling.
	\end{abstract}

	
	

	\section{Introduction} \label{intro}
	Massive stars are among the brightest and most luminous objects in the Galaxy. They host strong stellar winds that are essential for the chemical enrichment of their surrounding interstellar medium \citep{Abbott}.  
	
	The sample of known magnetic O-type stars in our Galaxy is currently very small. There are less than a dozen confirmed magnetic O-type stars \citep[e.g.][]{Wade2015a}. Inferred from spectropolarimetic observations, they are known to host strong (> 1\,kG), predominantly dipolar magnetic fields that are tilted with respect to the rotation axis of the star.  Consistent with the oblique magnetic rotator (OMR) paradigm, these stars often manifest periodic variability \citep[e.g.][]{Stahl1996}.

    A large fraction of the known Galactic magnetic O-type stars belong to the rare class of Of?p-type stars. Of?p-type stars are atypical Of stars that exhibit peculiar spectral properties. They were identified by the presence of C{\sc iii}$\lambda$4650 lines in emission that are in comparable strength to their neighbouring N{\sc iii} lines \citep{Walborn1972}. Of?p-type stars are also known to show recurrent spectral variability, notably in their Balmer and He I lines \citep[e.g.][]{Naze2001,Walborn2004,Naze2008}. As a result, Of?p-type stars can appear to change in spectral type based on their variable He{\sc ii} $\lambda$4541 to He{\sc i} $\lambda$4471 ratios. 

    The root cause behind the spectral peculiarities of the Of?p-type stars remains uncertain. However, stellar magnetism appears to be a common factor behind this phenomenon. Indeed, all known Galactic Of?p-type stars have been shown to be magnetic \citep[e.g.][]{Martins2010,Wade2011,Wade2012a,Wade2012b,Wade2015a,Grunhut2017}.    

    Massive magnetic stars are unique sites to observe the combined effect of stellar winds, rotation and magnetism. Understanding how stellar magnetism and rotation simultaneously affect the wind plasma on dynamical timescales requires a full magnetohydrodynamic (MHD) treatment. Such numerical calculations were performed by \citet{UDDoula1,UDDoula2,UDDoula3}. Their extensive MHD simulations show the temporal evolution of a wind-trapped magnetosphere. Coupling the MHD simulations with full radiative transfer can be used to predict the spectroscopic variability of magnetic massive stars. For instance, \citet{Sund2012} have successfully reproduced the H$\alpha$ equivalent width variability of HD 191612 using such a model. 

    A semi-analytical approach to magnetosphere modelling was first introduced by \citet{Town2005a}. Operating within the rigid-field limit, the magnetosphere resembles that of a corotating warped disk. This Rigidly Rotating Magnetosphere (RRM) model provided the theoretical basis for the construction of spectroscopic and photometric modelling tools.  \citet{Town2005b} and \citet{Town2008} found some success in using such a model to reproduce the spectroscopic and photometric variability of $\sigma$\,Ori\,E, a strongly magnetic Bp-type star.

    More recently has an Analytical Dynamical Magnetosphere (ADM) model been developed by \citet{Owocki2016}. ADM is specifically designed to characterize the dynamical magnetospheres of slowly rotating O-type stars. The magnetosphere structure that ADM produces is essentially an approximation of the time-averaged results of MHD simulations. Observations of HD 191612's  H$\alpha$ variability show that ADM is in good agreement with more sophisticated MHD simulations while being vastly more time efficient \citep{Owocki2016}. 

    Here, we present the first ADM-based photometric modelling tool capable of producing synthetic optical light curves of magnetic O-type stars. We exploit the ADM model as a means to simulate the density structure of their circumstellar magnetospheres. From this, we estimate the amount of light scattered via free electrons present in the wind plasma. Matching observations to models enables us to constrain important magnetic, stellar and wind parameters of magnetic massive stars. 

    The purpose of this investigation is thus to improve our general understanding of the magnetic properties of massive stars. As magnetic fields greatly affect their wind structure, characterizing the magnetic properties of O-type stars is of great interest.   

    The paper is organized as follows. In Section \ref{method}, we describe the light curve synthesis algorithm. We explore the parameters space of the model in Section \ref{parameters} and proceed with a validation test in Section \ref{validation}. This will be followed by two direct applications of our ADM-based photometric tool in Section \ref{applications}.  We conclude in the final section.

	\section{The numerical method} \label{method}
	
	We hypothesize that the photometric variability expected of a magnetic massive star primarily arises from the periodic occultation of an obliquely rotating envelope. In order to quantify these photometric fluctuations, we first need to characterise the magnetosphere through which we will perform radiative transfer. Both the magnetosphere model and radiative transfer model will be described in the following subsections.

        \subsection{Magnetosphere Model}

        We utilize the newly-developed ADM model to simulate the magnetosphere structure of slowly rotating O-type stars. ADM requires seven input parameters: the effective temperature ($T_\text{eff}$), stellar mass ($M_*)$, stellar radius $(R_*)$, terminal velocity ($v_\infty$), mass-loss rate\footnote{ The input mass-feeding rate ($\dot{M}_{B=0}$) is the hypothetical mass-loss rate of an unmagnetized star that is not to be confused with the actual mass-loss rate ($\dot{M}$). The presence of a dipolar field causes on overall reduction in the mass-loss rate (via magnetic confinement) such that $\dot{M} = f_B \dot{M}_{B=0}$ where $f_B \sim R_*/1.4 R_\text{A}$ \citep{UDDoula3} and $R_\text{A}$ is the Alfv\'en radius.} ($\dot{M}_{B=0}$), dipolar magnetic field strength ($B_\text{d}$) and smoothing length ($\delta$). 
    
        By default, ADM computes a 2D map of a magnetosphere that is aligned with the rotation axis. To simulate a misaligned dipole, two additional free parameters are required: the magnetic obliquity ($\beta$) and the inclination angle ($i$). The angle between the dipolar axis and the observer's line of sight ($\alpha$) may be expressed in terms of the $i$ and $\beta$ angles at each rotational phase ($\phi$):
    
        \begin{equation} \label{eq:alpha}
            \cos \alpha = \cos \beta \cos i +  \sin \beta \sin i \cos \phi , 
        \end{equation}
        where $\phi=0$ corresponds to a maximum on the light curve. At each rotational phase, the magnetosphere model must therefore be tilted by $\alpha$ before performing radiative transfer.

        \subsection{Radiative transfer}
        The periodic scattering of light caused by the presence of free electrons within the obliquely rotating envelope can be responsible for the photometric variability observed in magnetic massive stars. We can easily compute the amount of light attenuation in the single electron scattering limit ($\tau_e<1$, see below): 
        \begin{equation} \label{eq:light}
            I = I_0 e^{-\tau_e},
        \end{equation}
        where $I_0$ is the continuum intensity and $\tau_e$ is the electron scattering optical depth. 
    
        The electron scattering optical depth can be estimated from the total density. For core rays,
        \begin{equation} \label{eq:tau}
            \tau_e = \int_{R_*}^{\infty}  \frac{\alpha_e \sigma_e \rho}{m_\text{p}}  ds, 
        \end{equation}
        where $m_p$ is the proton mass, $\alpha_e$ is the number of free electrons per baryon mass, $\sigma_e$ is the electron scattering cross-section, $\rho$ is the magnetosphere mass density and $s$ is the path length between the stellar surface and the observer. For a completely ionized wind at solar metallicity, $\alpha_e \sim 0.85$. The lower boundary in eq. (\ref{eq:tau}), $R_*$, is defined as the onset of the wind. Non-core rays are not expected to contribute significantly in the single electron scattering limit and are therefore not considered in this analysis. 
    
        The flux is obtained by integrating the emergent intensity, eq. (\ref{eq:light}), over the occulted area of the star. In terms of differential magnitudes, the flux translates to
        \begin{equation} \label{eq:dm}
            \Delta m = \Delta m_0 + 2.5 \tau_e (\log e) ,
        \end{equation}
        where $\Delta m_0$ is a constant. Eq. (\ref{eq:dm}) is recalculated at each rotational phase in order to obtain a light curve. This is accomplished by first rotating the magnetosphere and then reevaluating eq. (\ref{eq:tau}). We can see that the amount of scattering material will be modulated by the magnetosphere tilt which is expected to be the cause of the  bulk of photometric variability.

        \subsection{Illustrative results}

    	\begin{figure*}
    	\includegraphics[width=\linewidth,trim={1.5cm 0.0cm 1.5cm 0.5cm},clip]{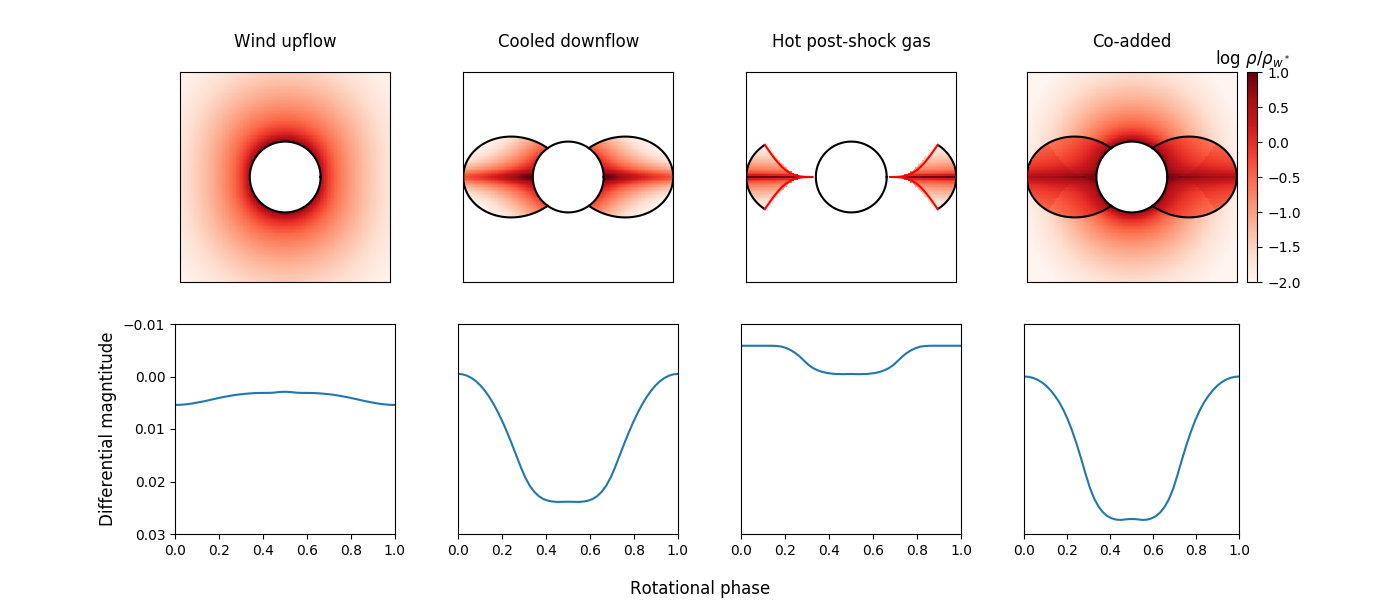}
    	\caption{Top: Density structure of the three ADM components (wind upflow, cooled downflow and hot post-shock) and the resulting co-added density. The magnetic axis is vertical and the densities are normalized to $\rho_{w^*}=\dot{M}_{B=0}/4\pi v_\infty R_*^2$. Bottom: Synthetic light curve corresponding to the different ADM components and the resulting light curve from the co-added density. }
    	\label{fig:ADM}
        \end{figure*}	
    
        \citet{Owocki2016} breaks down the structure of a dynamical magnetosphere into three distinct components: the wind outflow, the hot-post shock gas and the cooled downflow. Each component is co-added to create a 2D map of a magnetosphere. We rotate this map in order to generate a 3D data cube of an aligned magnetic dipole. 
    
        For illustration purposes, we show the density structure of the distinct ADM components in Fig. \ref{fig:ADM} (top row). The physical and magnetic parameters are similar to HD 191612, i.e. $T_\text{eff}=35$\,kK,  $M_*=30$\Msol, $R_*=15$\Rsol, $v_\infty=2700$\kms, $\dot{M}_{B=0}=10^{-6.0}$\Msol \yr and $B_\text{d}=2.5$\,kG  \citep{Wade2012a}, with $i=30^\circ$ and $\beta=60^\circ$. A smoothing length of $\delta/R_*=0.1$ was adopted.  
    
        We can see that the density of the wind upflow is only slightly asymmetric due to the presence of the magnetic field and tapers as $\sim 1/r$. Both the cooled downflow and  hot post-shock components are delimited by the Alfv\'en radius, $R_A \propto B_\text{eq}^{1/2}R_*^{1/2}/\dot{M}_{B=0}^{1/4} v_\infty^{1/4}$. Their equatorial regions represent high density areas. The amount of smoothing present within the cooled downflow component is controlled by  $\delta$.  Moreover, the size of the hot post-shock component is determined by the cooling parameter, $\chi_\infty \propto v_\infty^4 R_* / \dot{M}_{B=0}$. An increase in $\delta$ corresponds to enhanced mixing, while an increase in $\chi_\infty$ translates to an increase in shock radius \citep{Owocki2016}. 
    
        The computed light curves attributed to each of the density structure components are displayed in Fig. \ref{fig:ADM} (bottom row). It is apparent that the overall light curve is predominantly affected by the cooled downflow and hot post-shock components. They both cause periodic dimming in photometric brightness at $\phi=0.5$ (i.e. when the magnetosphere is viewed edge on to the observer). Conversely, the wind upflow component only weakly contributes to the resulting light curve and causes photometric attenuation at $\phi=0$. This is because, even in the presence of a dipolar magnetic field, the wind upflow component only becomes slightly more prolate, while the other two components are obviously oblate. 
    
        \citet{Henn2018} have recently demonstrated that the ADM model poorly reproduces the outer wind structure of real stars (by comparing UV resonance lines generated from sophisticated 3D radiative transfer simulations to the simplified ADM formulations). Since in our case the inner wind is the major photometric contributor, this is not an immediate concern.
    
\begin{figure*}
    	\includegraphics[width=\linewidth,trim={1.5cm 0.0cm 1.5cm 0.5cm},clip]{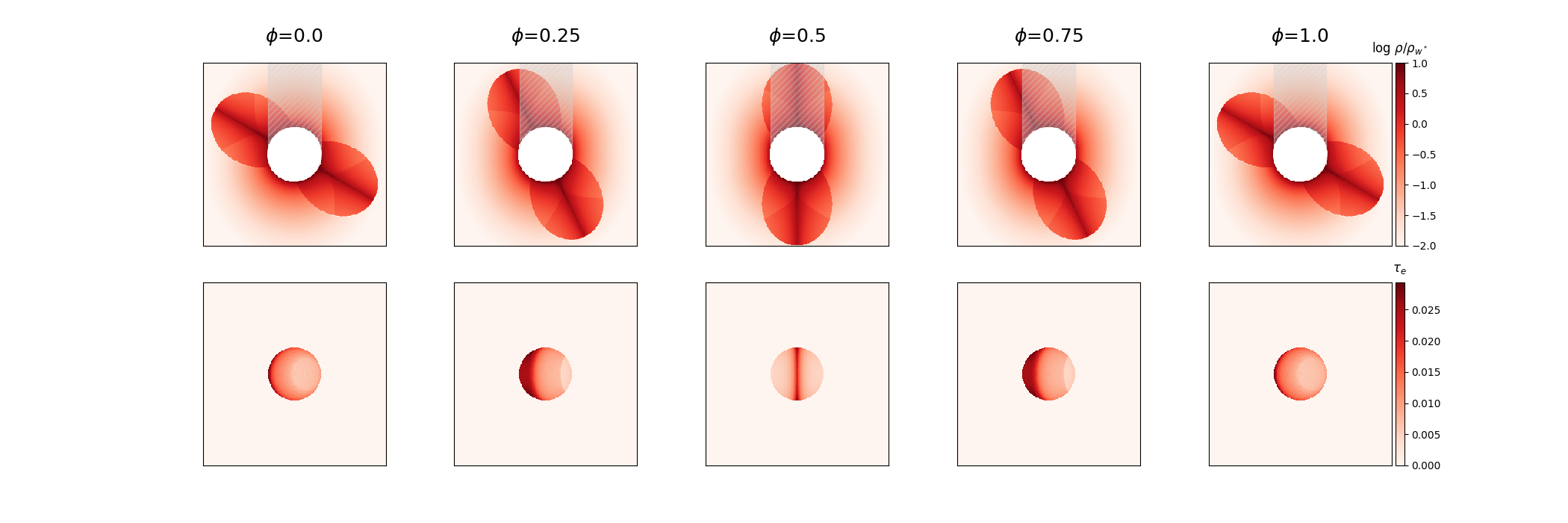}
    	\caption{
    	Top: 2D slice of the magnetosphere density structure. The density is normalized to $\rho_{w^*}=\dot{M}_{B=0}/4\pi v_\infty R_*^2$. 
    	The hashed region illustrates the part of magnetosphere that occults the star (as viewed from the observer's line-of-sight)}. Bottom: Optical depth along rays core and non-core rays. In all areas of the wind, the optical depth is below unity which validates the single scattering limit. 
        \label{fig:tau}
        \end{figure*}

        The light curves are computed within the optically thin limit ($\tau_e<1$). To justify this approximation, we display mosaics of the optical depth ($\tau_\infty$) along rays that intersect the stellar surface in the top panels of Fig. \ref{fig:tau}. The bottom panels of Fig. \ref{fig:tau} illustrate the tilt of the magnetosphere with respect to an observer arbitrarily placed along the z-axis. We notice that even when the optical depth is at its maximum, when the magnetic equator is seen edge-on (at $\phi=0.5$), the optical depth integrated over the complete ray is still well below unity.  This condition no longer holds in extreme cases of very large mass-loss rates (i.e. $\dot{M}>10^{-4.7}$ \Msol \yr), unphysically low terminal velocities (i.e. $v_\infty<25$\kms)  or high magnetic field strengths (i.e. $B_\text{d}>250$\,kG). Therefore, for typical massive star properties, the optically thin limit should remain valid.

        \subsection{Numerical Fitting}
        The rapid computation of magnetospheric light curves allows for the use of Markov chain Monte Carlo (MCMC) sampling methods when attempting to numerically fit models to observations.  We utilise a python implementation of MCMC - the \texttt{emcee} package developed by \citet{emcee}. 100 walkers were initialised with arbitrary initial parameters. The walkers were left to accomplish at least 1000 steps after burn-in. Burn-in is typically achieved between 100 and 500 steps. Depending on the noise level present within the observations, convergence may require more steps.  After convergence, confidence intervals were obtained from the likelihood distributions. 
    
	\section{Parameter space study} \label{parameters}
    
    The simplicity in our analytic models allows us to quickly compute grids of synthetic light curves. In order to understand the influence of the free parameters on the light curve synthesis model, we will conduct a parameter space study. 

    If we consider a star of fixed stellar parameters (i.e. $M_*$, $R_*$, $T_\text{eff}$), the remaining free parameters are the geometric angles, the wind parameters, the magnetic field strength, and the smoothing length. We will therefore explore the error space of the following six free parameters: $i$,\,$\beta$,\, $\dot{M}_{B=0}$,\,$B_\text{d}$,\,$v_\infty$ and $\delta$. 

    Figs. \ref{fig:Mdot}, \ref{fig:Bd}, \ref{fig:vinf} and  \ref{fig:delta} display grids of synthetic light curves as a function of $i$ and $\beta$ among the set $i, \beta=\{10,\,30,\,50,\,70\}^\circ$. Each grid shows the linear increase of one free parameter encoded in ADM on a light curve with basic parameters  $T_\text{eff}=35$\,kK,  $M_*=30$\Msol, $R_*=15$\Rsol, $v_\infty=2500$\kms, $\dot{M}_{B=0}=10^{-6.0}$\Msol \yr, $B_\text{d}=2.5$\,kG and $\delta = 0.1 R_*$. In each grid, the free parameter respectively ranges from $\dot{M}_{B=0} = \{1.0,\,2.0,\,3.0\}\times 10^{-6}$\,\Msol\yr, $B_\text{d}=\{2.5,\,5.0,\,7.5\}$\,kG, $v_\infty=\{1.5,\,2.5,\,3.5\} \times 1000$\,\kms and $\delta=\{0.1,\,0.3,\,0.5\}\,R_*$. Their effect on the overall light curve shape is discussed below.

        \subsection{The geometric angles}
        
        The $i$ and $\beta$ angles control both the shape and amplitude of the photometric modulations. In fact, for $i + \beta < 90^\circ$, the light curves are single-dipped, while they become double-dipped for $i + \beta > 90^\circ$. As the curves transition from single- to double-dipped at $i + \beta = 90^\circ$, the curves become flat-bottomed. The amplitude of the variability increases as $i$ (or $\beta$) approaches $90^\circ$. Conversely, as $i$ (or $\beta$) tends to $0^\circ$, the variability decreases.   
    
        The dips in the photometry are caused by an increase in optical depth when the magnetic equator crosses the observer's line-of-sight. This occurs twice during a rotational cycle if $i + \beta > 90^\circ$ (such that both magnetic hemispheres can be seen by the observer). Two dips in the light curve are expected to occur at phases $\phi_\text{dip1} = \arccos \left( - \cot i \cot \beta \right)  $ and $\phi_\text{dip2} = 1 - \phi_\text{dip1} $, corresponding to a dip separation of $\Delta \phi_\text{dips} = 1 - 2 \phi_\text{dip1}$.  If $i + \beta < 90^\circ$, the magnetic equator does not quite cross the observer's line-of-sight, which will instead only cause one dip to appear in the light curve. This occurs when the magnetic equator is closest to the observer's line of sight at phase $\phi_\text{dip} = 0.5$).

        Furthermore, according to eq. (\ref{eq:alpha}), we can see that the $i$ and $\beta$ angles are interchangeable and thus degenerate. This explains the symmetry in the grid of light curves along the diagonal.  To avoid this degeneracy numerically, eq. (\ref{eq:alpha}) can be re-expressed as a function of two new variables that are no longer degenerate, $i+\beta$ and  $|i-\beta|$, such that 
        \begin{equation} \label{eq:alpha2}
            \cos\alpha=\frac{1}{2}\cos(\beta-i)\left[1+\cos\phi\right]+\frac{1}{2}\cos(\beta+i)\left[1-\cos\phi\right].
        \end{equation}
        As a result, with photometric data alone, it is only possible to uniquely determine the sum of the $i$ and $\beta$ angles and the absolute value of their difference. From this we can obtain two possible solutions of $(i,\beta)$ couples that are degenerate and indistinguishable photometrically.
        
        \subsection{The mass-feeding rate}
\begin{figure*}
    	\includegraphics[width=0.62\linewidth,trim={0 0.25cm 0 1.75cm},clip]{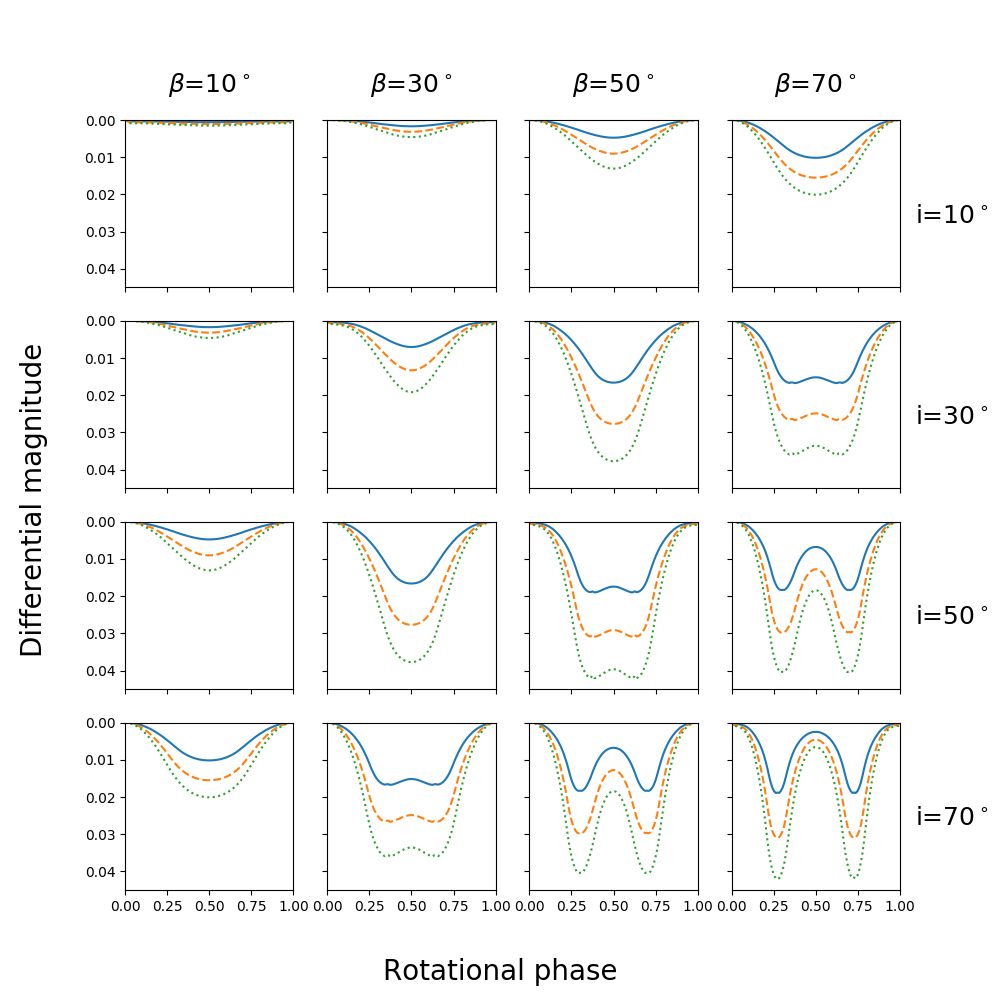}
    	\caption{Grid of model light curves as a function of inclination, $i$,  and obliquity, $\beta$. Overplotted are curves of varying mass-feeding rate,$\dot{M}_{B=0}$. The solid (blue), dashed (orange) and dotted (green) lines respectively correspond to values of $\dot{M}_{B=0}=\{1.0,2.0,3.0\}\times10^{-6}$\Msol\yr. The remaining stellar, wind or magnetic values are based on HD 191612. Note that the grid is symmetric along the diagonal. This illustrates the degeneracy between the $i$ and $\beta$ angles.}
    	\label{fig:Mdot}
        \end{figure*}

        As shown in Fig. \ref{fig:Mdot},  the depth of the light curve minima appear to scale linearly with the mass-feeding rate. This is an expected result in the optically thin limit as 1) the differential magnitude is proportional to the opacity, 2) the opacity scales with the density and 3) the density  is is proportional to the mass-feeding rate ($\rho\propto\dot{M}_{B=0} $). We note that the mass-feeding rate can also have a minor impact on the sizes of the magnetosphere radius ($R_A \propto \dot{M}_{B=0}^{-1/4}$) and the hot-post shock region ($\chi_\infty \propto \dot{M}_{B=0}^{-1}$). However, these two consequences negate each other so that the resulting light depth shape remains roughly linear with the mass-feeding rate due to the density scaling. Since electron scattering is a $\rho$-dependent process (optically thin), wind clumping does not play an important role. Furthermore, optically thick clumping (macro-clumping or porosity) does not need to be considered, due to $\tau_e < 1$.
    
        \subsection{The dipolar magnetic field strength}
\begin{figure*}
    	\includegraphics[width=0.62\linewidth,trim={0 0.25cm 0 1.75cm},clip]{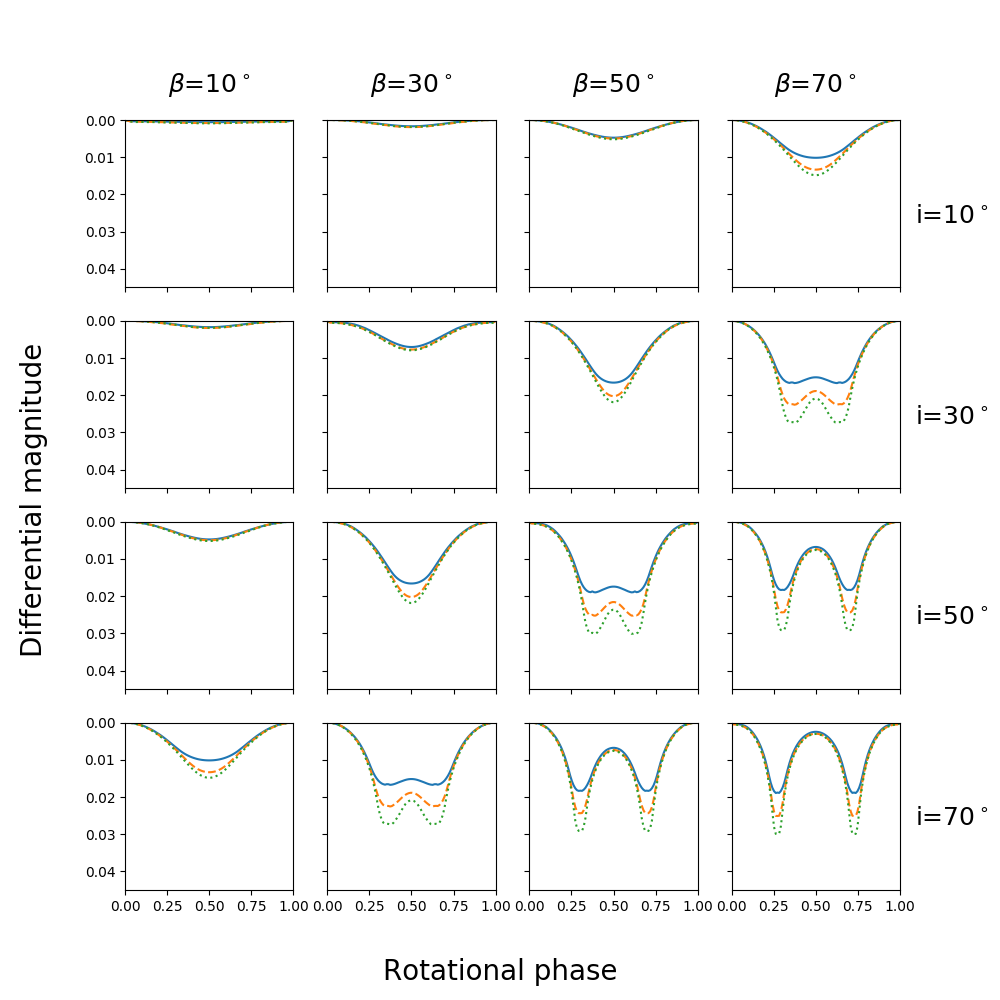}
    	\caption{Same as Fig. \ref{fig:Mdot} for varying dipolar field strength, $B_\text{d}$. The solid (blue), dashed (orange) and dotted (green) lines respectively correspond to values of  $B_\text{d}=\{2.5,5.0,7.5\}$\,kG.  }
    	\label{fig:Bd}
        \end{figure*}

        The dipolar field strength plays another important role in the light curve occultation depth. Similar to the mass-feeding rate, an increase in $B_\text{d}$ yields an increase in occultation depth (see Fig. \ref{fig:Bd}). $\dot{M}_{B=0}$ and $B_\text{d}$ are therefore somewhat degenerate with respect to each other. However, unlike the mass-feeding rate the dipolar field strength does not scale linearly  with the light curve depth. For instance, the effect of increasing $B_\text{d}$ becomes less important at large $B_\text{d}$ values. In addition, obvious changes in the light curve shape can be perceived as the dipolar field strength increases. Notably, when $i+\beta>90^\circ$, the magnitude depth of the maximum between the two dips seems to increase along with $B_\text{d}$. This occurs as $B_\text{d}$ primarily only affects the magnetosphere radius ($R_A \propto B_\text{d}^{1/2}$). 
        
        \subsection{The terminal velocity}
\begin{figure*}
    	\includegraphics[width=0.62\linewidth,trim={0 0.25cm 0 1.75cm},clip]{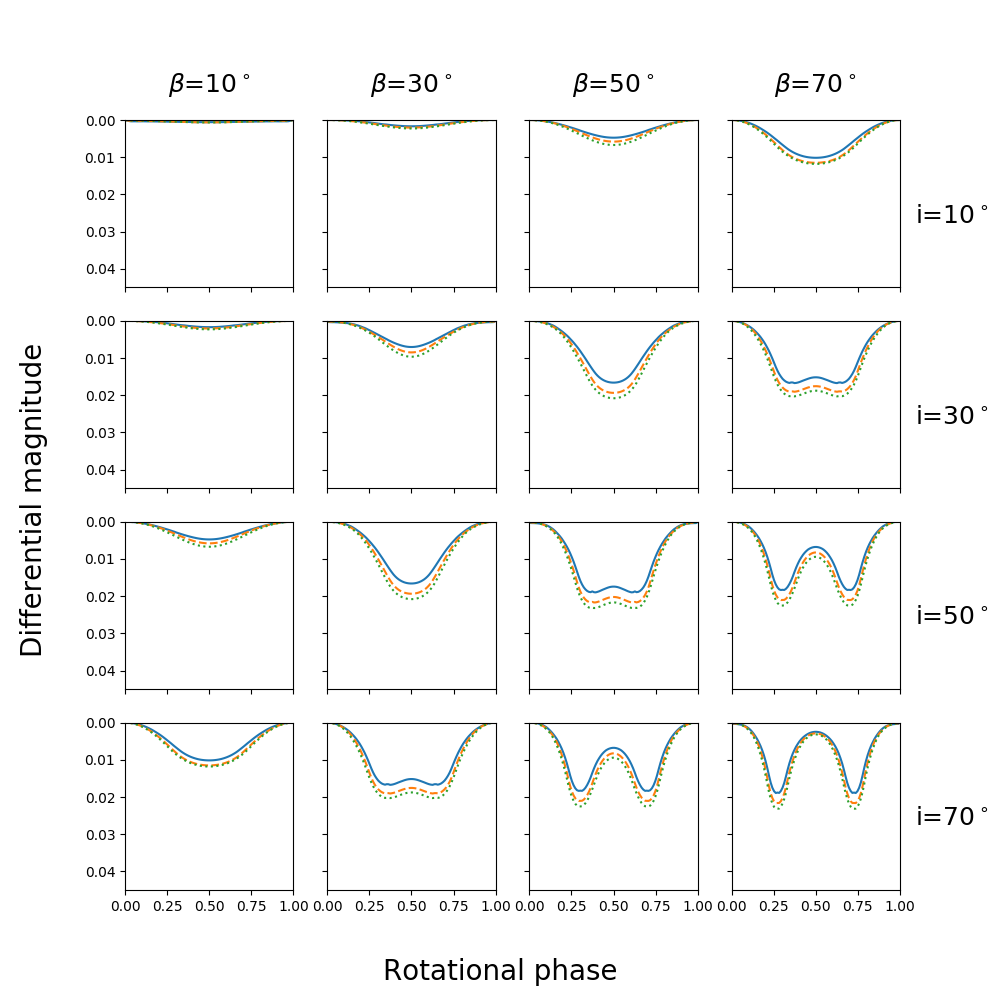}
    	\caption{Same as Fig. \ref{fig:Mdot} for varying terminal velocity, $v_\infty$. The solid (blue), dashed (orange) and dotted (green) lines respectively correspond to values of  $v_\infty=\{1500,2500,3500\}$\kms.}
    	\label{fig:vinf}
        \end{figure*}

        According to Fig. \ref{fig:vinf}, we notice that the terminal velocity only has a minor effect on the total light curve depth. This is not a surprising result in the wind upflow and cooled downflow components since the Alfv\'en radius is only weakly dependent on the terminal velocity ($R_A \propto v_\infty^{-1/4}$), whereas in the hot post-shock region, the shock radius is highly dependent on the terminal velocity ($\chi_\infty \propto v_\infty^4$). However, this effect is partially counteracted by its dependence on the overall magnetosphere density ($\rho \propto v_\infty^{-1}$). Overall, an increase in terminal velocity effectively leads to a small increase in light curve occultation depth primarily originating from the hot post-shock component. 

         \subsection{The smoothing length}
\begin{figure*}
    	\includegraphics[width=0.62\linewidth,trim={0 0.25cm 0 1.75cm},clip]{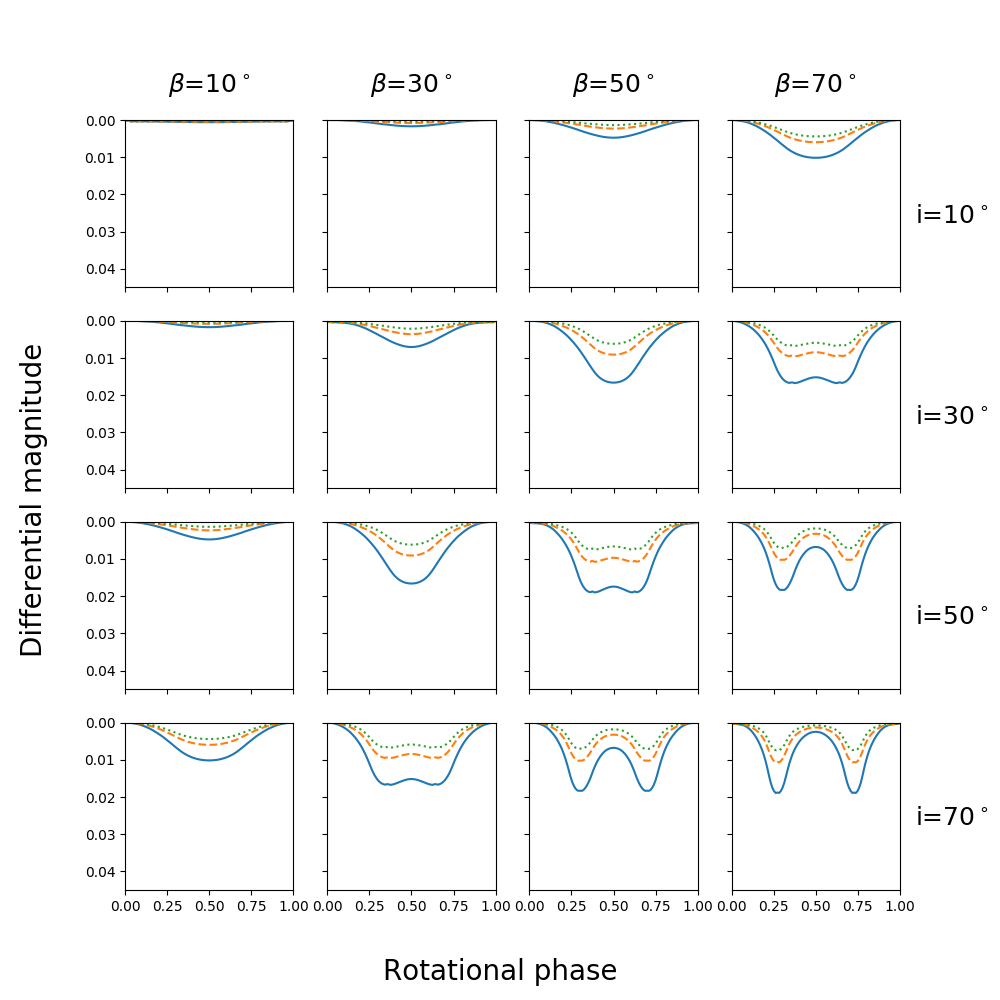}
    	\caption{Same as Fig. \ref{fig:Mdot} for varying smoothing length, $\delta$. The solid (blue), dashed (orange) and dotted (green) lines respectively correspond to values of  $\delta/R_*=\{0.1, 0.3, 0.5\}$.}
    	\label{fig:delta}
        \end{figure*}

        The smoothing length only affects the cooled downflow component in ADM. An increase in $\delta$ promotes an increase in spatial smoothing and thus a decrease in light curve depth (see Fig. \ref{fig:delta}). Among all the other free parameters forming the ADM model, $\delta$ is the most uncertain. However, for HD 191612, it is suspected to range between $0.1-0.5\,R_*$ \citep{Owocki2016}. To more reliably constrain the smoothing length, it would be instructive to match observations to theoretical predictions where the other wind and magnetic parameters, such as the mass-loss rate and dipole magnetic field strength, are already well known. For the rest of this paper, $\delta/R_*=0.1$ was adopted.

    \section{Numerical simulations} \label{validation}
    A key validation of our approach lines in the ability of our photometric model to retrieve the ``correct" stellar or magnetic parameters, especially when realistic noise is present in the data. Indeed, in a case where a signal is buried in noise, numerous input parameters may produce seemingly adequate fits. How well does our ADM-based diagnostic tool perform?
	
	Ultimately, our goal is to model the photometric signatures of the suspected magnetic Of?p-type stars of the Magellanic Clouds as a means to predict their magnetic field strengths and magnetosphere geometry. In anticipation of this, we will first model synthetic light curves to which we have added Gaussian noise that mimics the dispersion present within the OGLE observations of Magellanic Of?p-type stars \citep{Naze2015}. Obtaining parameters of best-fit from the simulated noise models and comparing them to the known parameters of the initial model light curves will enable us to validate our fitting routine.

	We estimate the OGLE noise by obtaining flat residuals to the observed photometry after the subtraction of an appropriate harmonic fit. The residuals obey a Gaussian-like distribution about a central mean. We obtain a dispersion of $\sigma_\text{OGLE}=8$ mmag that is consistent across the OGLE sample of Magellanic Of?p-type stars. 
	
\begin{figure*}
	\includegraphics[width=\linewidth]{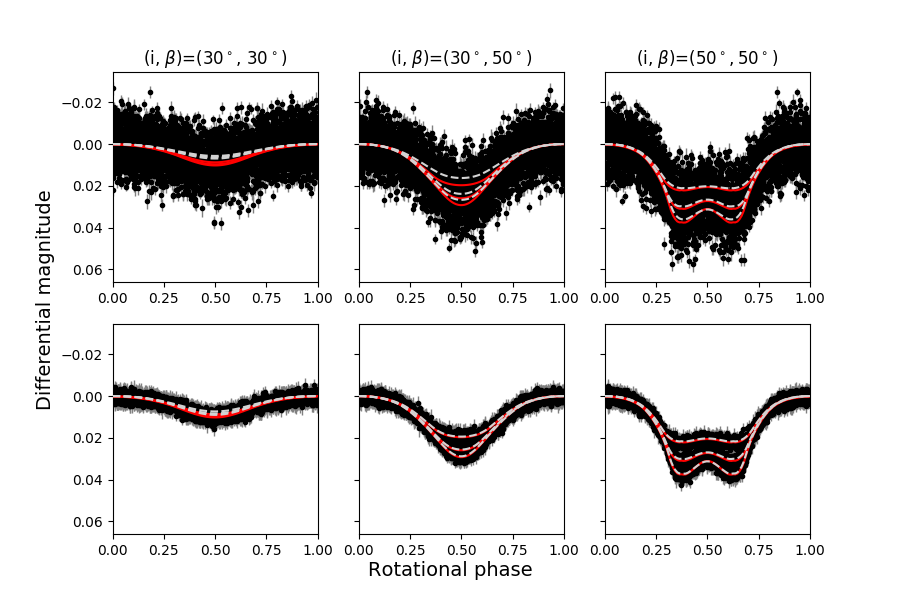}
	\caption{Synthetic light curves with OGLE noise (top) and reduced OGLE noise (bottom) added. The inclination and obliquity angles are increasing from left to right: $(i,\beta)=\{(30^\circ,30^\circ),(30^\circ,50^\circ), (50^\circ,50^\circ)\}$. In each panel, curves of increasing magnetic field strength are overplotted. The solid (red) lines correspond to the model light curve models before artificial noise is added. The dashed (gray) lines correspond to best-fit light curves to the noisy models. }
	\label{fig:noise}
    \end{figure*}

	The set of synthetic light curves that we will consider have input parameters: $(i,\beta)=\{(30^\circ,30^\circ),(30^\circ,50^\circ), (50^\circ,50^\circ)\}$ and $B_\text{d}=\{2.5,5.0,7.5\}$\,kG. The grid of models with simulated OGLE noise are illustrated in Fig. \ref{fig:noise}. The simulated noise models were binned at every 0.05 phase for a total of 20 bins. We utilize Monte Carlo Markov Chains methods on the binned light-curves to obtain the curves of best-fit. Table \ref{tab:noise} shows the comparison of the true model parameters and the obtained best-fit parameters.	The noise light curves are shown in Fig. \ref{fig:noise} (top row).

	We find that at low inclination and obliquity angles, our model struggles to retrieve the true input parameters, notably in the $|i-\beta|$ angle. However, in all cases, the sum $i+\beta$ was determined accurately. Unfortunately, uncertainty in $|i-\beta|$ will propagate into increased error in both the  inclination and obliquity angles, even if $i+\beta$ is precise. Furthermore, success in retrieving the correct magnetic field strength also increased at higher inclination and obliquity angles. This is to be expected as the ADM model becomes more sensitive to $B_\text{d}$ at increased $i+\beta$ angles. The confidence intervals for $B_\text{d}$ generally yield upper bound uncertainties that are larger than the lower bound uncertainties. This is a consequence of two model characteristics: the insensitivity of our model at large $B_\text{d}$ values (as it primarily only affects the equatorial density) and the degeneracy between $B_\text{d}$ and the couple $(i,\beta)$ (i.e. an increase in magnetic field strength can be compensated by a decrease in inclination and/or obliquity angles).

	In a more optimistic approach, we also consider models with reduced noise. The added noise is one-fifth of the OGLE dispersion: $\sigma_\text{OGLE}/5$. These synthetic light curves are displayed in Fig. \ref{fig:noise} (bottom row). The best-fit parameters are listed in Table \ref{tab:noise}. As expected, with reduced noise added, the ability of our photometric model to obtain the true input parameters was increased. Similar trends can be noted from the previous case: light curves with higher inclination and obliquity angles are more accurately reproduced, notwithstanding the reduced noise. 

	\begin{table*}
	\centering
	\caption{Comparison between input model parameters and obtained best-fit parameters.}
	\label{tab:noise}
	\begin{tabular}{ccccccccccccc} 
		\hline
		\multicolumn{3}{c}{Model} & \multicolumn{5}{c}{Noise $\sigma_\text{OGLE}$} & \multicolumn{5}{c}{Noise $\sigma_\text{OGLE}/5$} \\
		\multicolumn{3}{c}{Input parameters} & \multicolumn{5}{c}{best-fit parameters} & \multicolumn{5}{c}{best-fit parameters} \\ \cmidrule(lr){1-3} \cmidrule(lr){4-8} \cmidrule(lr){9-13}
		$ B_\text{d}$   & $i$  & $\beta$ & $B_\text{d}$& $i+\beta$ & $|i-\beta|$ & $i$ or $\beta$ & $i$ or $\beta$  & $B_\text{d}$ & $i+\beta$ & $|i-\beta|$ & $i$ or $\beta$ & $i$ or $\beta$ \\
		$[\text{kG}]$ & [deg]& [deg] & [kG] & [deg]      & [deg]       & [deg]          & [deg]         & [kG]   & [deg]      & [deg]       & [deg]          & [deg]          \\
		\hline
		\multirow{3}{*}{$2.5$}  & $30$& $30$  & $3.2^{+2.5}_{-1.0}$  & $55^{+8}_{-6}$    &$19^{+24}_{-13}$&$17^{+8}_{-9}$&$37^{+15}_{-8}$  & $4.5^{+3.4}_{-3.0}$& $60^{+4}_{-2}$  &$10^{+10}_{-7}$ &$25^{+4}_{-5}$ &$36^{+6}_{-4}$  \\
		& $30$& $50$   & $2.8^{+3.5}_{-1.2}$ & $76^{+5}_{-3}$   &$28^{+23}_{-19}$&$24^{+9}_{-10}$&$52^{+12}_{-9}$& $2.3^{+0.3}_{-0.2}$& $79.7^{+0.7}_{-0.7}$ &$16^{+10}_{-10}$&$32^{+5}_{-5}$ &$48^{+5}_{-5}$  \\
		& $50$& $50$    & $2.3^{+0.1}_{-0.1}$& $98^{+1}_{-1}$    &$9^{+9}_{-6}$&$44^{+3}_{-5}$&$54^{+4}_{-3}$ & $2.5^{+0.1}_{-0.1}$& $99.7^{+0.5}_{-0.6}$ &$8^{+6}_{-5}$   &$46^{+3}_{-3}$ &$54^{+3}_{-3}$  \\
		\multirow{3}{*}{$5.0$}    & $30$& $30$    & $4.0^{+3.2}_{-1.0}$& $56^{+5}_{-4}$    &$13^{18}_{-9}$&$21^{+5}_{-8}$&$34^{+10}_{-5}$ & $6.4^{+4.9}_{-4.0}$& $60^{+2}_{-1}$       &$8^{+9}_{-6}$   &$26^{+3}_{-4}$ &$34^{+5}_{-3}$  \\
		& $30$& $50$   & $4.6^{+3.3}_{-1.1}$ & $79^{+3}_{-2}$    &$20^{+20}_{-14}$&$29^{+7}_{-10}$&$50^{+10}_{-7}$ & $5.2^{+1.6}_{-0.6}$& $80.0^{+0.5}_{-0.6}$ &$20^{+6}_{-8}$  &$30^{+4}_{-3}$ &$50^{+3}_{-4}$  \\
		& $50$& $50$   & $4.7^{+0.2}_{-0.2}$ & $98^{+3}_{-4}$    &$25^{+21}_{-17}$&$36^{+9}_{-12}$&$62^{+9}_{-8}$ & $5.0^{+0.1}_{-0.1}$& $99.6^{+0.3}_{-0.4}$ &$11^{+6}_{-7}$  &$44^{+4}_{-3}$ &$55^{+3}_{-3}$  \\
		\multirow{3}{*}{$7.5$}   & $30$& $30$    & $6.0^{+5.2}_{-4.8}$& $62^{+8}_{-6}$    &$27^{+25}_{-19}$&$16^{+9}_{-9}$&$44^{+17}_{-11}$ & $7.9^{+4.9}_{-4.7}$& $60^{+2}_{-1}$       &$11^{+10}_{-7}$ &$25^{+3}_{-5}$ &$36^{+5}_{-4}$  \\
		& $30$& $50$    & $6.9^{+4.2}_{-1.1}$& $80^{+2}_{-2}$   &$24^{+20}_{-16}$&$28^{+8}_{-10}$&$51^{+10}_{-8}$& $7.1^{+1.4}_{-0.8}$& $79.9^{+0.6}_{-0.7}$ &$23^{+5}_{-9}$  &$28^{+4}_{-3}$ &$51^{+3}_{-4}$  \\
		& $50$& $50$    & $7.4^{+0.4}_{-0.4}$& $99^{+1}_{-1}$    &$5^{+5}_{-4}$&$47^{+2}_{-3}$&$52^{+3}_{-2}$ & $7.5^{+0.1}_{-0.1}$& $99.6^{+0.1}_{-0.1}$ &$9^{+5}_{-6}$ &$45^{+3}_{-3}$ &$54^{+2}_{-3}$  \\
		\hline
	\end{tabular}
\end{table*}

	\section{Applications} \label{applications}
    The photometric model that we have just described can in principle be used as a diagnostic tool to determine the wind and magnetic properties of magnetic O-type stars. We will demonstrate two direct applications of our model as a magnetic field strength determination tool. Such a capability is valuable in cases where direct spectropolarimetric detections of magnetic fields are unfeasible, as, for example, has so far been the case for the Magellanic Of?p-type stars.

    	\subsection{HD 191612}
        HD 191612 is the prototypical Of?p type star, known to display repeatable photometric and spectroscopic variability which was first reported by \cite{Walborn2004}. \citet{Donati2006} obtained the first direct magnetic detection of this star that led to the suspicion that HD 191612 was in fact a magnetic star exhibiting rotational modulation. Since then, the photometric, spectroscopic and magnetic variability of HD 191612 have been shown to vary in phase over the same $\sim 537$\,d  thus confirming the oblique magnetic rotator paradigm of HD 191612 \citep{Wade2011}.

        The photometric data of HD 191612 were obtained from Hipparcos observations. They were phase folded according to the period and ephemeris provided by \citet{Wade2011}. We fix HD 191612's stellar and wind parameters according to the most recent revisions by \citet{Wade2011} and \citet{Howarth2007}\footnote{\citet{Howarth2007} had derived a clumped mass-loss rate of $10^{-5.8}$ \Msol \yr via H$\alpha$ diagnostics. Using a clumping factor of 4 and a magnetic reduction factor of 0.25, we obtain a mass-feeding rate of $10^{-5.8}$ \Msol \yr. Alternatively, utilizing \citet{Vink2001} recipes, we obtain an unclumped mass-feeding rate of $10^{-6}$ \Msol \yr based on their observational results. These results are consistent within error. For this experiment, we have adopted the larger value of $\dot{M}_{B=0} = 10^{-5.8}$ \Msol \yr. }:  $T_\text{eff}=35$\,kK, $M_*=30$\Msol, $R_*=15$\Rsol, $v_\infty=2700$\kms and $\dot{M}_{B=0}=10^{-5.8}$\Msol \yr.  Since we expect to be able to use our model as a magnetic field strength determination tool,  $B_\text{d}$, $i$ and $\beta$ will be free parameters, and all of the others will be fixed.  

\begin{figure}
        	\includegraphics[width=\columnwidth]{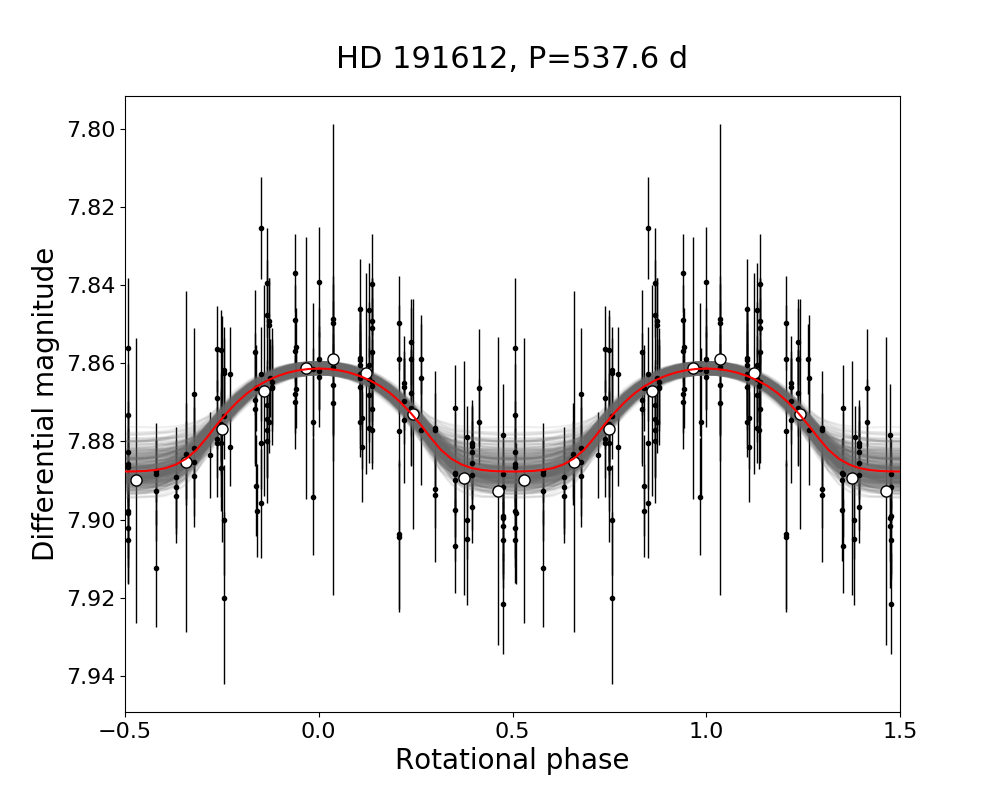}
        	\caption{Phased raw (filled circles) and binned (open circles)  Hipparcos light curve for HD 191612. The curve of best-fit is overplotted in red (bold solid curves). Sample curves that span the 1 $\sigma$ error bars on the best-fit parameters are overplotted in gray (thin solid lines).}
        	\label{fig:fit}
        \end{figure}

        To avoid degeneracy between $i$ and $\beta$, we use $i+\beta$ and $|i-\beta|$ as free parameters (see eq. (\ref{eq:alpha2})). We derive the couple of ($i,\beta$) solutions afterwards.  The curve of best-fit to the binned light curve is showed in Fig. \ref{fig:fit} and the parameters of best-fit are listed in Table \ref{tab:fit}. The likelihood distributions for the model parameters are shown in Fig. \ref{fig:dist}

        Our model reproduces the photometric observations fairly well. Though the individual angles cannot be uniquely determined, their sum, yielding $i+\beta=88_{-5}^{+5}$\,$^\circ$, is well constrained. This value is consistent with previous tentative results from \citet{Wade2011}, who had obtained $i+\beta\sim95\pm10^\circ$ by modelling the longitudinal magnetic field strength of HD 191612. Despite the scatter present in the Hipparcos light curve, we obtain a dipole field strength of $B_\text{d}=2.7^{+0.6}_{-0.4}$\,kG. This agrees with the Zeeman-inferred value of $B_\text{d}\sim2.5\pm0.4$\,kG by \citet{Wade2011}.  We note that the majority of the uncertainty derives from the $|i-\beta|$ angle. This could be mitigated with improved photometric data.  

        \begin{table}
        	\centering
        	\caption{Best-fit parameters to the Hipparcos photometry of HD 191612}
        	\label{tab:fit}
        	\begin{tabular}{ccccccc} 
        		\hline
        		Star &$i+\beta$&$|i-\beta|$&  $i$ or $\beta$ & $i$ or $\beta$ & $B_\text{d}$ & $\Delta m_0$ \\
        		        &[deg]&  [deg]  &[deg]&[deg]&[kG]&[mmag]\\
        		\hline
        		HD 191612&$88_{-5}^{+5}$& $33_{-23}^{+26}$& $27_{-14}^{+13}$& $61_{-11}^{+13}$ & $2.7_{-0.4}^{+0.6}$ & $7861_{-2}^{+2}$ \\
        		\hline
        	\end{tabular}
        \end{table}

	\subsection{The Of?p stars of the Magellanic Clouds}
	There are six known Of?p-type stars in the Galaxy:  HD  108,  HD  148937,  HD  191612,  NGC 1624-2, CPD-28$^\circ$ 256 and $\theta^1$\,Ori\,C \citep[e.g.][]{Walborn2000,Jesus2019}. Magnetic fields have been firmly detected in each of these stars \citep[see][]{Martins2010,Wade2011,Wade2012a,Wade2012b,Wade2015a}.  They were shown to display distinctive periodic variations across numerous observable quantities, notably in their photometric brightness and spectral line strength, that complies with the oblique magnetic rotator model. 

	The stellar and magnetic properties of the known Galactic Of?p-type stars have been thoroughly studied. They are known to be widely diverse; in fact, their rotational periods range from days to years and their magnetic fields range from 1 to 20\,kG. In order to draw more definite conclusions about the enigmatic  Of?p class of stars, a larger sample is required. 

	A small sample of Of?p-type stars has recently been detected in the Small and Large Magellanic Clouds. Presented in chronological order of detection, they are: Av 200, 2dFS 936, BI 57, LMC164-2, SMC159-2 and LMCe136-1 \citep[see][]{Walborn2000,Massey2001,Massey2014,Neugent2018}. Parallel spectroscopic and photometric studies of these stars were led by \citet{Walborn2015} and \citet{Naze2015} respectively. The photometric data were obtained from the Optical Gravitational Lensing Experiment \citep[OGLE,][]{OGLE}. In most cases, \citet{Naze2015} were successful in identifying a single dominating period that characterizes the photometric variability of each star. Meanwhile, \citet{Walborn2015} had obtained additional spectroscopic observations at the Las Campanas Observatory with  the  Boller  \&  Chivens  spectrograph attached to the 2.5 m du Pont telescope. \citet{Walborn2015} reported spectroscopic variability, analogous to that of Galactic Of?p stars, that is in phase with the photometric variations for the majority of the Magellanic Of?p-type stars.

	Because of the similarity with Galactic Of?p stars, the Of?p-type stars of the Magellanic Clouds are considered to be the first candidate extra-Galactic magnetic O-type stars. However, magnetic fields have yet to be detected in these stars. Attempts by \citet{Bagnulo2017} were made to measure their magnetic fields via circular spectropolarization. Unfortunately, only marginal or null detections were obtained (see also Bagnulo et al. in prep). Modelling their photometric variability can therefore provide an alternative, indirect method of estimating the magnetic field strengths of the Magellanic Of?p-type stars. 

    We therefore employed our ADM-based modelling tool to fit the phased light curves of these stars. To fit for $i$, $\beta$ and $B_\text{d}$, we fix $T_\text{eff}$, $M_*$, $R_*$, $v_\infty$ and $\dot{M}_{B=0}$. Effective temperatures and surface gravities were obtained from line-profile fitting to synthetic hydrogen and helium spectra calculated by means of the NLTE, unified model atmosphere/spectrum synthesis code {\sc FASTWIND} \citep[V10.1, see][and references therein]{Gonzalez2012}. In particular, we compared with a precalculated grid of models (adopting a LMC metallicity $Z=0.5$\,$Z_\odot$, and a SMC metallicity $Z=0.2$\,$Z_\odot$), with stepsize $\Delta T_\text{eff}$ = 1000 K and $\Delta \log g$ = 0.1 dex. For the mass-loss rates, we adopted values prototypical for the considered objects, and fixed the helium content and the micro-turbulent velocity to 10\% of hydrogen (by number) and 10 kms$^{-1}$, respectively. In most cases (but see below), we obtained reasonable fits to the hydrogen line wings ($\log g$) and the He{\sc ii} and He{\sc i} lines ($T_\text{eff}$), leaving the peculiar He{\sc
    II}\,4686 aside. Due to noisy and contaminated spectra, the uncertainties for effective temperature and gravity are somewhat larger than for high S/N spectra of non-magnetic stars, and particularly the gravities for LMC164-2 and SMC159-2 remain rather uncertain. The remaining parameters where subsequently derived from a combination of $T_\text{eff}$, $\log g$ and observed quantities, as described in the notes to Table \ref{tab:MC}. Their wind properties were scaled according to the metallicities of the Small and Large Magellanic Clouds using \citet{Leitherer1992} and \citet{Vink2001} recipes.  Table \ref{tab:MC} summarizes their stellar and wind parameters along with their rotational periods. their spectral types were obtained at the minimum emission stage. 
    
    The OGLE light curves of the Of?p-type stars in the Magellanic Clouds were phased according to their respective periods and ephemerides that were provided by \citet{Naze2015} and Bagnulo et al. (in prep). In \citet{Naze2015}, $\phi=0$ corresponds to a photometric minima, instead of a photometric maximum that was adopted in Section \ref{method}. We therefore rephased the light curves to agree with our convention. In most cases, this corresponds to a phase shift of 0.5 (e.g. 2dFS 936, SMC159-2, LMC164-2). For Bi 57 and LMCe136-1 (where the curves are either double-waved or asymmetric), a phase shift of 0.57 and 0.6 was arbitrarily adopted to localise a global minima near 0.5 phase.
    
\begin{figure*}
	\includegraphics[width=\linewidth, trim={0 10cm 0 1.5cm}, clip]{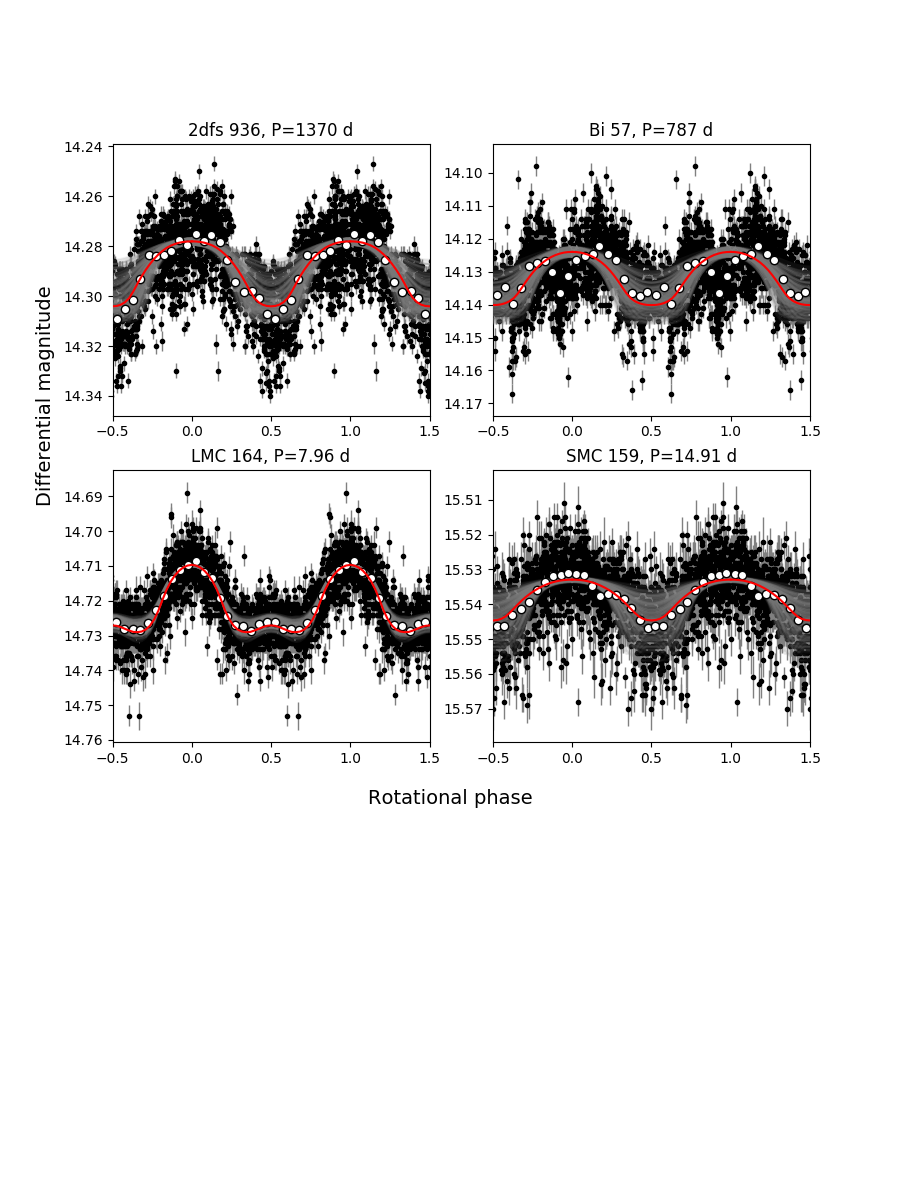}
	\caption{Phased raw (filled circles) and binned (open circles) OGLE light curves of 2dFS 936, BI 57, LMC164-2, SMC159-2. The curve of best-fit is overplotted in red (bold solid lines). Curves that span the 1 $\sigma$ error bars on the best-fit parameters are overplotted in gray (thin solid lines).}
	\label{fig:fitOGLE1}
\end{figure*}

\begin{figure}
	\includegraphics[width=\linewidth, trim={3.0 1.5cm 11cm 18cm}, clip]{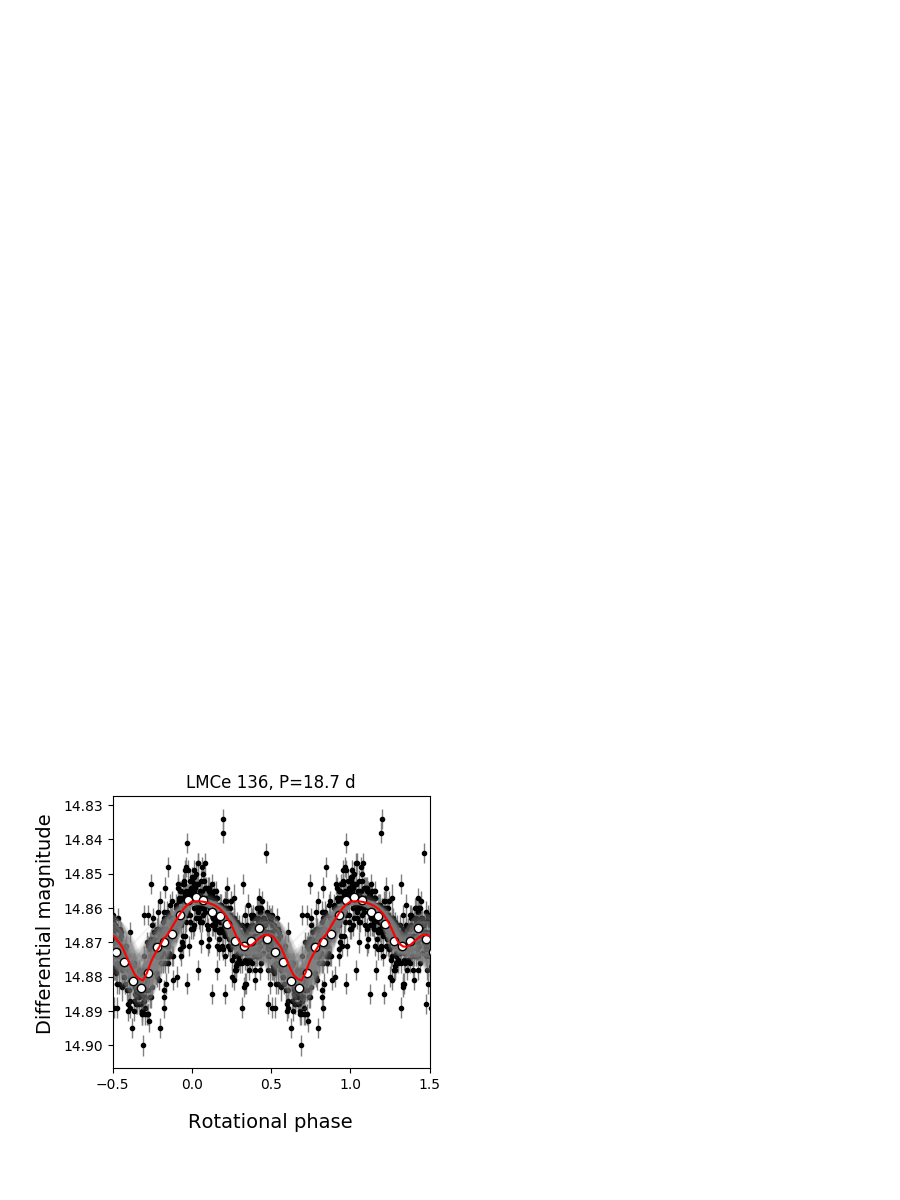}
	\caption{Phased raw (filled circles) and binned (open circles) OGLE light curve of LMCe136-1. The curve of best-fit is overplotted in red (bold solid lines). Curves that span the 1 $\sigma$ error bars on the best-fit parameters are overplotted in gray (thin solid lines).}
	\label{fig:fitOGLE2}
\end{figure}

    Prior to fitting, the OGLE light curves were binned in phase (20 bins). The best-fit curves are displayed in Figs. \ref{fig:fitOGLE1} and \ref{fig:fitOGLE2}. The corresponding best-fit parameters are listed in tables \ref{tab:fit1} and \ref{tab:fit2}. Figs. \ref{fig:dist1} to \ref{fig:dist5} respectively show the likelihood distributions of the model parameters for 2dFS 936, Bi 57, LMC164-2,  SMC159-2 and LMCe136-1. AV 220 does not have a well-defined period and was therefore omitted from this analysis.  Details on the remaining individuals stars are elaborated bellow.

	\begin{table*}
	\begin{center}
	\caption{Summary of the properties of known Of?p-type stars from the Small and Large Magellanic Clouds. }
	\begin{tabular}{l c c c c c c c c c c}
			\hline
	         &	Spectral	   &T$_\text{eff}^\text{a}$ & $\log$\,g$^\text{a}$	&$M_\text{v}^\text{b,c}$&$\log L_*^\text{d}$ 	&$R_*^\text{e}$		   &	$M_*^\text{f}$	&$v_\infty^\text{g}$   &$\log\dot{M}_{B=0}^\text{h}$&$P_\text{rot}$	\\
	         & Type		   & 	[kK]	   &	[cgs]	&            &[\Lsol]	    &[\Rsol]       &[\Msol]     &	[km/s]    &[\Msol\yr]   & [d]	        \\
			\hline
	AV 220	 &O5.5 f?p &40$\pm$1.5	   &4.0$\pm$0.15 &$-4.9^b$    &5.32$\pm$0.03	&10.1$\pm$0.3  &29.9$\pm$8	&2330$\pm$300 &-6.5$\pm$0.2	& $>1000^\text{i}$	\\	 
	2dFS 936 &O6.5 f?p &42$\pm$1.5	   &4.0$\pm$0.15 &$-5.6^b$    &5.65$\pm$0.03	&13.4$\pm$0.4  &51.3$\pm$14 &2640$\pm$300 &-6.0$\pm$0.2	&$1370\pm30^\text{i}$		\\
	BI 57	 &O7.5 f?p &36$\pm$1.5	   &3.7$\pm$0.15 &$-5.0^b$    &5.23$\pm$0.03	&11.3$\pm$0.4  &22.4$\pm$5	&1850$\pm$200 &-6.2$\pm$0.2	&$787\pm14^\text{i}$		\\
	LMC164-2 &O8 f?p   &40$\pm$1.5	   &4.0$\pm$0.15 &$-4.5^c$    &5.16$\pm$0.03	&8.4$\pm$0.3   &20.7$\pm$6	&2330$\pm$300 &-6.4$\pm$0.2	&$7.959\pm0.003^\text{i}$\\
	SMC159-2 &O8 f?p   &40$\pm$1.5	   &4.0$\pm$0.25 &$-4.0^c$    &4.96$\pm$0.03	&6.6$\pm$0.2   &13.1$\pm$4	&1900$\pm$200 &-6.8$\pm$0.2	&$14.914\pm0.004^\text{i}$ 		\\
	LMCe136-1 &O6.5 f?p   &40$\pm$1.5	   &4.05$\pm$0.25 &$-4.3^c$   &5.10$\pm$0.03&7.5$\pm$0.2   &23.0$\pm$4	&2170$\pm$200 &-6.5$\pm$0.2	&$18.706\pm0.016^\text{j}$\\
			\hline	
		\end{tabular}

	\begin{tablenotes}
      \small
      \item $^\text{a}$ Effective temperatures and surface gravities were obtained in this paper.
      \item $^\text{b,c}$ Absolute magnitudes were obtained from \citet{Walborn2015} and \citet{Massey2014} respectively.
      \item $^\text{d}$ Luminosities were derived using absolute magnitudes and  theoretical bolometric corrections from \citet{Martins2010} and visual magnitudes.
      \item $^\text{e}$ Stellar radii were derived from the effective temperature and luminosity according to Stefan-Boltzmann law. 
      \item $^\text{f}$ Stellar masses were derived from the surface gravity and stellar radius. 
      \item $^\text{g}$ Terminal velocities were determined using the \citet{Leitherer1992} scaling relation.
      \item $^\text{h}$ Mass-loss rates were determined from the \citet{Vink2001} recipes. 
      \item $^\text{i,j}$ Rotational periods are those reported by \citet{Naze2015} and Bagnulo et al. (in prep) respectively. 
    \end{tablenotes}
	\label{tab:MC}
	\end{center}
	\end{table*}

        \subsubsection{2dFS 936 in the SMC}
        In their search for Wolf-Rayet stars in the SMC, 2dFS 936 was identified as an Of?p-type stars by \citet{Massey2001}. Its spectral type ranges from O6.5 to O7. \citet{Naze2015} derived a photometric period of $1370\pm30$ d. Spectroscopic data obtained by \citet{Walborn2015} agreed with this period. Among the known extra-Galactic Of?p-type stars, 2dFS 936 exhibits the largest depth of photometric variability. Its light curve can be reproduced with $(i,\beta)=(\beta,i)=(18_{-8}^{+13},65_{-11}^{+8})^\circ$ and $B_\text{d}= 7.2_{-1.3}^{+1.8}$\,kG. We note that the variability present in the OGLE light curve is sharper than the modelled light curve. This is not a limitation of the model, but rather a consequence of scatter present in the observed photometry. Prior to fitting, we bin the light curve which smooths out variability and can thus lead to a flattened best-fit model.

        \subsubsection{BI 57 in the LMC}
        BI 57 is an O7.5 to O8 star and a member of the LMC. Two periods were detected in the OGLE photometry by \citet{Naze2015}: a $400\pm3.5$ d and a $787\pm14$ d period. From the photometric observations alone it is not possible to distinguish which period represents the rotational period of the star. EW measurements reported by \citet{Walborn2015} and \citet{Bagnulo2017} appear to confirm the longer $787\pm14$ d period. Modelling the photometric variability (phase folded according to the $\sim787$ d period) yields $(i,\beta)=(\beta,i)=(25_{-12}^{+11},56_{-10}^{+14})^\circ$ and $B_\text{d}=2.3_{-1.0}^{+1.9}$\,kG. However, there are obvious discrepancies between this theoretical light curve and the observations. Indeed, there are two photometric minima during one rotational cycle (i.e. double-wave) which is not expected of our model.

        \subsubsection{LMC164-2 in the LMC}
        LMC164-2 was identified as an Of?p-type star during a more extensive search for WR stars in both the SMC and LMC by \citet{Massey2014}. At minimum emission phases, LMC164-2 has the characteristics of an O8 spectral type. The photometric study by \citet{Naze2015} extracted a period of $7.9606\pm0.0010$ d. This is the shortest period detected in the sample of extra-Galactic Of?p-type stars. Among the Magellanic Of?p-type stars, LMC164-2 has the highest inferred dipolar field strength. A double-dipped minimum can be seen in the light curve, which strongly constrains the $i+\beta$ angle. We obtain a magnetic field strength of $12.3_{-0.5}^{+1.8}$\,kG and a magnetic geometry of $(i,\beta)=(\beta,i)=(11_{-1}^{+3},85_{-1}^{+1})^\circ$.  The best-fit light curve appears to accurately describe the photometric variability.

        \subsubsection{SMC159-2 in the SMC}
        SMC159-2 is an O8?p-type star in the SMC. It was also detected within the framework of the WR survey in the Magellanic Clouds by \citep{Massey2014}. \citet{Naze2015} detected a period $14.914\pm0.004$ d from the OGLE photometry. The curve of best-fit to the observations yields  $(i,\beta)=(\beta,i)=(16_{-9}^{+14},67_{-4}^{+10})^\circ$ and $B_\text{d}=6.4_{-1.8}^{+3.5}$\,kG. This star's magnetic field strength is not as reliably constrained as those of the other Magellanic Of?p-type stars. Indeed, while the lower limit is quite well defined, the likelihood distribution present a long tail towards large dipole field strengths (see Fig. \ref{fig:dist4}). Similar to the case of 2dFS 936, the photometry of SMC159-2 also suffers from a large dispersion. Fitting only to the binned light curve, the best-fit model does not perfectly fit the light curve near $\phi=0$ and $\phi=1$.

        \subsubsection{LMCe136-1 in the LMC}
        LMCe136-1 is the most recent addition to the sample of Magellanic Cloud Of?p-type stars. This new Of?p-type star was discovered by \citet{Neugent2018} in the LMC. Its spectrum is consistent with that of an O6.5f?p-type star. A period of $18.914\pm0.004$ d was detected from the OGLE photometry (Bagnulo et al. in prep). The light curve of LMCe136-1 is asymmetric: two dips occur with different magnitude depths. This feature cannot be explained by a centered dipolar field topology. Instead, we consider an offset dipole model. Preliminary results recognize a significant dipole offset of $a = 0.21^{+0.03}_{-0.03}\,R_*$ (perpendicular to the magnetic axis) with basic parameters $i\sim57_{-19}^{+21}$\,$^\circ$, $\beta=51_{-21}^{+20}$\,$^\circ$ and $B_\text{d}=5.6_{-1.9}^{+4.1}$\,kG. A model light curve with these optimised parameters can adequately reproduce the observed photometry. Further details on the implementation of an offset-dipole magnetic field topology with ADM are described in Section \ref{offset}.

\begin{table}
	\begin{center}
	\caption{Best-fit parameters to the OGLE photometry with a dipole model}
	\begin{tabular}{cccccccc} 
		\hline
		Star    &$i+\beta$      &$|i-\beta|$      & $i$ or $\beta$  & $\beta$ or $i$    & $B_\text{d}$ &$\Delta m_0$ \\
		        &[deg]          &[deg]            & [deg]           &    [deg]          &[kG]                   &   [mmag] \\
		\hline
		2dFS 936&$84_{-3}^{+3}$&$47_{-23}^{+17}$& $18_{-8}^{+13}$  & $65_{-11}^{+8}$  & $7.2_{-1.3}^{+1.8}$  &$14278_{-1}^{+1}$ \\
		BI 57    &$83_{-9}^{+6}$&$29_{-20}^{+26}$& $25_{-12}^{+11}$ & $56_{-10}^{+14}$ & $2.3_{-1.0}^{+1.9}$  &$14123_{-2}^{+2}$    \\
		LMC164-2 &$95_{-1}^{+1}$&$74_{-4}^{+2}$  & $11_{-1}^{+3}$   & $85_{-1}^{+1}$   & $12.3_{-0.5}^{+1.8}$ &$14711_{-1}^{+1}$ \\
		SMC159-2 &$83_{-2}^{+3}$&$50_{-28}^{+18}$& $16_{-9}^{+14}$  & $67_{-4}^{+10}$  & $6.4_{-1.8}^{+3.5}$  &$15533_{-1}^{+1}$ \\
		\hline
	\end{tabular}
    \label{tab:fit1}
    \end{center}
\end{table}

\begin{table}
	\begin{center}
	\caption{Best-fit parameters to the OGLE photometry with an offset-dipole model}
	\begin{tabular}{ccccccc} 
		\hline
		Star       & $i$  & $\beta$ & $B_\text{d}$ & $\Delta m_0$ & $a$\\
		           & [deg]           &    [deg]          &[kG]                   &   [mmag] & [\Rsol] \\
		\hline
		LMCe136-1 &   $57_{-19}^{+21}$ & $51_{-21}^{+20}$ & $5.6_{-1.9}^{+4.1}$  &$14858_{-1}^{+1}$ & $a=-0.21^{+0.03}_{-0.03}$ \\
		\hline
	\end{tabular}
    \label{tab:fit2}
    \end{center}
\end{table}

	\section{Discussion} \label{discussion}
	
		\subsection{General agreement of model fits to the observations}
	
    	In most cases, the photometric data can be reasonably well reproduced by a model light curve and we have provided estimates on the corresponding magnetic field strength and geometry. LMC164-2 and SMC159-2 fit particularly well with a dipole model. The $i+\beta$ angles could be determined accurately, however, uncertainty in the $i-\beta$ value was consistently large. The dipolar field strengths were generally well constrained with smaller lower bound limits but larger upper bound uncertainties. However, we admit that our best-fit results are sensitive to the (fixed) value of the mass-fed rate and to the (fixed) value of the smoothing length. For instance, $B_\text{d}$ will most likely be underestimated if $\dot{M}_{B=0}$ is overestimated or overestimated if $\dot{M}_{B=0}$ is underestimated. Similarly, adopting a low smoothing length, like in our case, may result in overestimations of the magnetic field strength. 
    	
    	In some cases, the meaningfulness of the best-fit light curves is more questionable. Both 2dFS 936 and BI 57 present some level of irregular variability that is quite possibly unrelated to their rotational modulations. For 2dFS 936, some additional signal may be coming from a blended eclipsing binary \citep[see][]{Naze2015}. For BI 57, it is also evident that magnetospheric electron scattering cannot explain the full range of photometric variability present in the OGLE data.  
    
        LMCe136-1, the latest addition to the Magellanic Of?p-type stars, stands out among the others of its kind. Though coherent, its photometric variability is noticeably asymmetric (i.e. the light curve between phase 0 and 0.5 does not mirror the curve between phase 0.5 and 1.0) which we interpret as due to deviations from a centered dipole model. This is interesting from both a theoretical and observational standpoint. While decentred dipoles are commonplace for Ap and Bp type stars \citep[for offsets parallel to the magnetic field axis][]{Mathys1997}, solid evidence supporting this geometry is currently lacking for magnetic O-type stars. LMCe136-1 thus provides a unique opportunity to explore and possibly even confirm more exotic magnetic field topologies. As demonstrated, LMCe136-1 can be more appropriately modeled with the implementation of an offset dipole model. Our tentative decentered dipole modelling efforts agree rather well with the observations.

	    \subsection{Offset-dipole magnetic field topologies} \label{offset}

        The asymmetry present in the light curve of LMCe136-1 motivated the expansion of ADM to an offset-dipole paradigm. Indeed, a dipole offset (perpendicular to the magnetic axis) will cause an increase in optical depth on one edge of the magnetosphere, yet a decrease in optical depth in the other, thus causing the overall light curve to appear asymmetric. More specifically, for configurations with $i+\beta<90^\circ$, the single dip will be skewed, while for for configurations with $i+\beta>90^\circ$, the double dips will be uneven. 
    
        The implementation of an offset-dipole magnetic field topology with ADM is relatively trivial. For simplicity, only offsets along the magnetic axis are considered. This introduces a new free parameter, the dipole offset ($a$). After the 3D magnetosphere is computed (see Section \ref{method}), the hot-post shock and cooled downflow components are shifted along the magnetic field axis according to the dipole offset parameter. Because of the loss of axial symmetry, eq. ($\ref{eq:alpha}$) no longer holds as a means to readily tilt the magnetosphere with respect to the observers line-of-sight. The magnetosphere must therefore be tilted sequentially. This is accomplished by: 1) performing a rotation by angle $\beta$ to simulate a misaligned dipole, 2) rotating the entire magnetosphere with respect to the rotation axis of the star by an angle equivalent to the rotational phase, 3) carrying out a rotation by angle $i$ to tilt the rotated magnetosphere with respect to the observer's line-of-sight. 
    
        In this decentered dipole case, the $i$ and $\beta$ angles are explicitly fit instead of being derived from their sum and difference (as described in the centered dipole case). The offset dipole paradigm therefore involuntarily relieves the degeneracy between the inclination and obliquity angles. Explicit details on the effect of a decentred dipole on the photometric and polarimetric features of magnetic O-type stars will be discussed in a future paper. 
	
	\subsection{Stellar magnetism and metallicity}

	Until recently, massive magnetic stars have only been observed in the vicinity of the solar neighbourhood. The newly-discovered extra-Galactic candidate magnetic stars have opened the possibility of studying stellar magnetism at different metal compositions. 
	
	Metallicity has a direct impact on the wind momentum of massive stars. This in turn will have an important effect on the wind characteristics of a star (i.e. terminal velocity and mass-loss rate). According to \citet{Leitherer1992} and \citet{Vink2001}, both the terminal velocity and mass-loss rate are expected to scale with the metallicity as $v_\infty \sim Z^{0.13}$ and $\dot{M}_{B=0} \sim Z^{0.69}$. 

	The Small and Large Magellanic Clouds are known to have significantly lower metallicities compared to the Galaxy. In fact, for the SMC and LMC respectively, $Z_\text{SMC} = 0.2$ $Z_\odot$ and $Z_\text{LMC} = 0.5$ $Z_\odot$ \citep[e.g.][]{ZMC}. As a result, the Of?p-type stars of the Magellanic Clouds are expected to have reduced mass-loss rates according to their respective birthplace metallicities. 
    
    The OGLE photometry for these stars have revealed large occultation depths that are comparable to that of the Galactic Of?p type star HD 191612. This may seem like an unexpected result considering the significantly lowered mass-loss rates of the Magellanic Of?p-type stars. As a consequence, in order to reproduce their light curve depths these stars are inferred to have systematically stronger magnetic fields than the magnetic O stars in the Galaxy. Indeed, the magnetic field strengths that we have derived for these stars are all consistently higher than the typical Galactic magnetic O-type star. This is especially noticeable for LMC164-2 ($B_\text{d} \sim 12$\,kG). Alternatively, this could be result of slower leakage from the magnetosphere.

	\subsection{Longitudinal magnetic field predictions}
\begin{figure*}
	\includegraphics[width=\linewidth,trim={0cm 0.1cm 0cm 0.cm},clip]{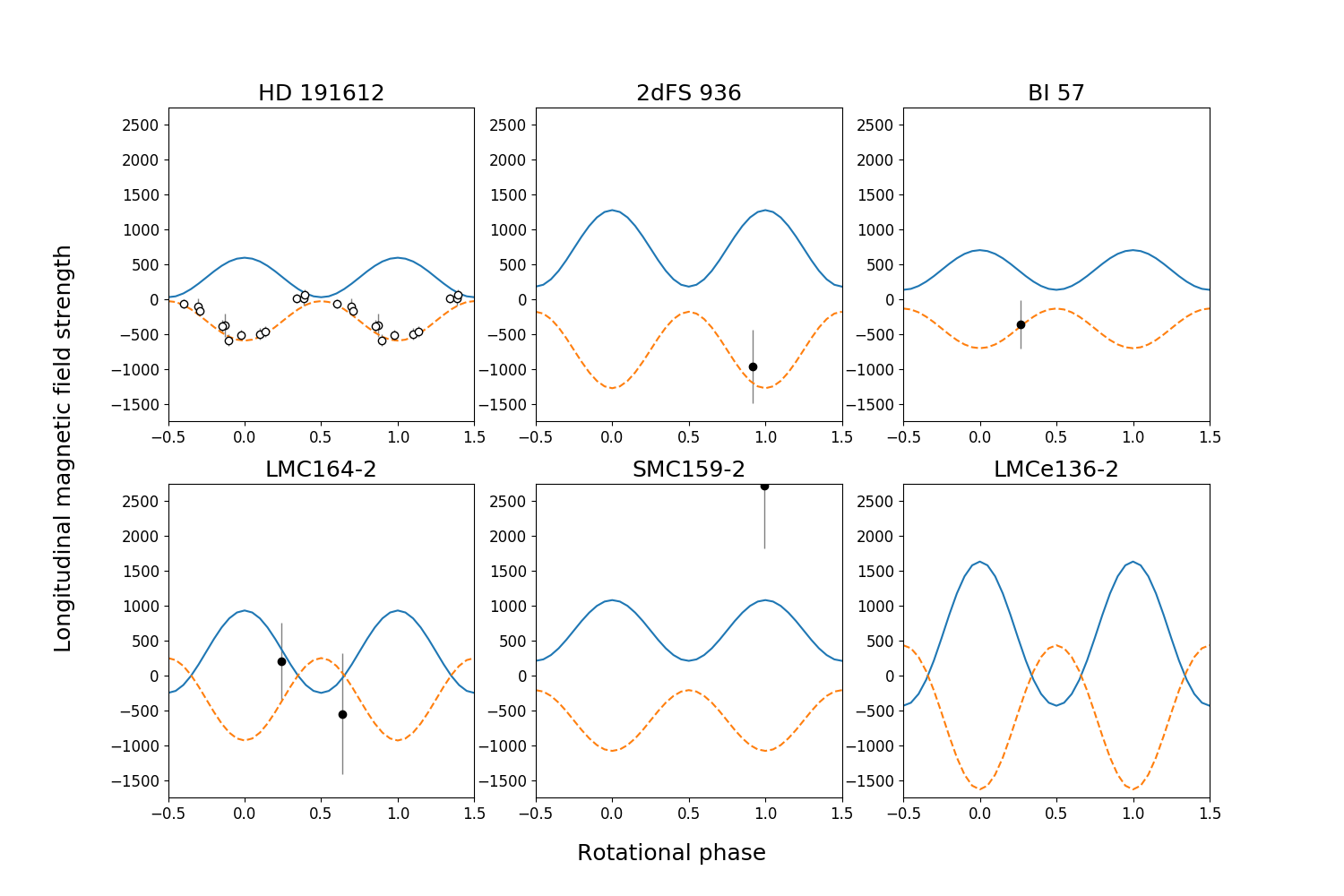}
	\caption{Model longitudinal magnetic field strength predictions of HD 191612 and five of the Magellanic Of?p-type stars: 2dFS 936, BI 57, LMC164-2, SMC159-2 and LMCe136-1. The blue solid curves correspond to $+B_z$ and the orange dashed curves to $-B_z$. Overploted are longitudinal magnetic field measurements from \citet{Wade2012a} (open circles), \citet{Bagnulo2017} (filled circles). The data points were rephased according to our convention of $\phi$ (i.e. photometric maximum at $\phi=0$).}
	\label{fig:Bzall}
    \end{figure*}

	Utilizing our inferred dipolar field strengths and geometries, we can predict the longitudinal magnetic field strength ($B_z$) variability of the Magellanic Of?p-type stars. We calculate $B_z$ using to the FLDCURV program written by J.D. Landstreet. However, since the photometric modelling is insensitive to the sign of the magnetic poles, we do not have any information on which pole of the magnetosphere is viewed as the poles are symmetric in mass structure above the equator. This leads to two possible solutions in the longitudinal magnetic field strength curves, $\pm B_z$, that only observations may be able to constrain. Both the positive (solid blue curve) and negative (dashed orange curve) solutions for $B_z$ are shown in Fig. \ref{fig:Bzall} for HD 191612 and the sample of Of?p-type stars of the Magellanic Clouds. The predicted longitudinal magnetic field strength curves were calculated with a limb darkening factor of 0.3. 
	
	 For HD 191612, the measured longitudinal magnetic field strengths were obtained by \citet{Wade2011} from Stokes V spectra.  The $-B_z$ curve is in good agreement with the observations.  
	
	Spectropolarimetric measurements of the extra-Galactic Of?p-type stars were carried out by \citet{Bagnulo2017} and Bagnulo et al. (in prep) with FORS2. No magnetic fields were firmly detected. However, the observed data points appear to be loosely consistent with at least one of the predicted longitudinal magnetic field curves. In some cases (e.g. 2dFS 936 and SMC159-2), the observations may even be able to distinguish the $\pm B_z$ curves as they appear to favour one solution over another. Typical error bars ranged from 0.4 to 1.0\,kG, with the except of LMCe136 where the error bars are roughly 0.3\,kG. In order to achieve the detection threshold for FORS2 ($5\sigma$), errors bars less than 0.3\,kG are required. LMCe136-1 is nominally predicted to be the best candidate for detection if it is observed at a favorable phase. 
	
	\section{Conclusion} \label{conclu}
	Magnetospheric single electron scattering can be responsible for the photometric signatures of magnetic O-type stars. In this paper, we have derived a light curve synthesis tool that can reproduce the photometric variability of an oblique magnetic rotator. 

    Our algorithm relies on the Analytical Dynamical Magnetosphere (ADM) model developed by \citet{Owocki2016} as a means to simulate the electron scattering magnetospheres of magnetic O-type stars. From this, we estimate the electron density structure and treat it within the single electron scattering regime. By integrating over the entire column density that is occulting the star (at numerous rotational phases), we obtain a theoretical light curve of a magnetized O-type star. 
    
    By characterizing the photometric variability perceived from magnetic massive stars, we can gain some useful insight about the fundamental processes that occur within their magnetospheres. This is particularly useful for parameterizing their magnetically confined winds. Our ADM-based photometric tool can therefore be used as a means to determine the mass-loss rates, the magnetic field strengths, or affiliated geometric angles (i.e. $i$ and $\beta$) of magnetic massive stars.

    We have conducted a pseudo-blind test on HD 191612, a well-behaved prototypical Of?p-type star with known stellar and magnetic parameters. We fix HD 191612's stellar and wind parameters and match a theoretical light curve to its Hipparcos photometry. We recover a magnetic geometry ($i+\beta\sim88_{-5}^{+5}$\,$^\circ$) and a field strength ($B_\text{d}=2.7^{+0.6}_{-0.4}$\,kG) that are consistent with previous investigations, yet obtained independently. This supports the utility of our model as a magnetic field strength diagnostic tool. 
 
	Next, we have applied the ADM-based magnetic field strength diagnostic tool to the the Of?p type stars of Magellanic Clouds. These stars are highly suspected to host magnetic fields, however spectropolarimetric observations have yet to detect the magnitudes of their fields. Interpreting their photometric variability within the framework of an Oblique Magnetic Rotator paradigm, we have derived a magnetic field strength and geometry (see Table \ref{tab:MC}) that is compatible with their observations, and presented predictions of their expected longitudinal magnetic field variations.
	
	Finally, we have presented the first preliminary results of a decentered dipole model with ADM. This extention to ADM was important to model the asymmetric light curve of LMCe136-1. We note that asymmetries similar to the photometric light curve of LMCe136-1, were seen in the $H\alpha$ EW variations of $\theta^1$\,Ori\,C \citep{Stahl1996}. We suspect that this feature is also indicative of deviations from a centered dipole model but from a spectroscopic perspective. This highlights the importance of observing numerous observable quantities and attempting to simultaneously fit them.

	\section*{Acknowledgements}
	GAW acknowledges support from a Natural Sciences and Engineering Research Council (NSERC) of Canada Discovery Grant. Y.N. acknowledges support from the Fonds National de la Recherche Scientifique (Belgium), the Communaut\'e Fran\c caise de Belgique, the European Space Agency (ESA) and the Belgian Federal Science Policy Office (BELSPO) in the framework of the PRODEX Programme (contract XMaS). M.K.Sz. acknowledges support from the Polish National Science Centre, grant MAESTRO2014/14/A/ST9/00121.
	
	
	
	
	\bibliographystyle{mnras}
	\bibliography{example} 

	
	
	
	\appendix

	\section{Likelihood distributions}
	
\begin{figure}
	\includegraphics[width=\columnwidth]{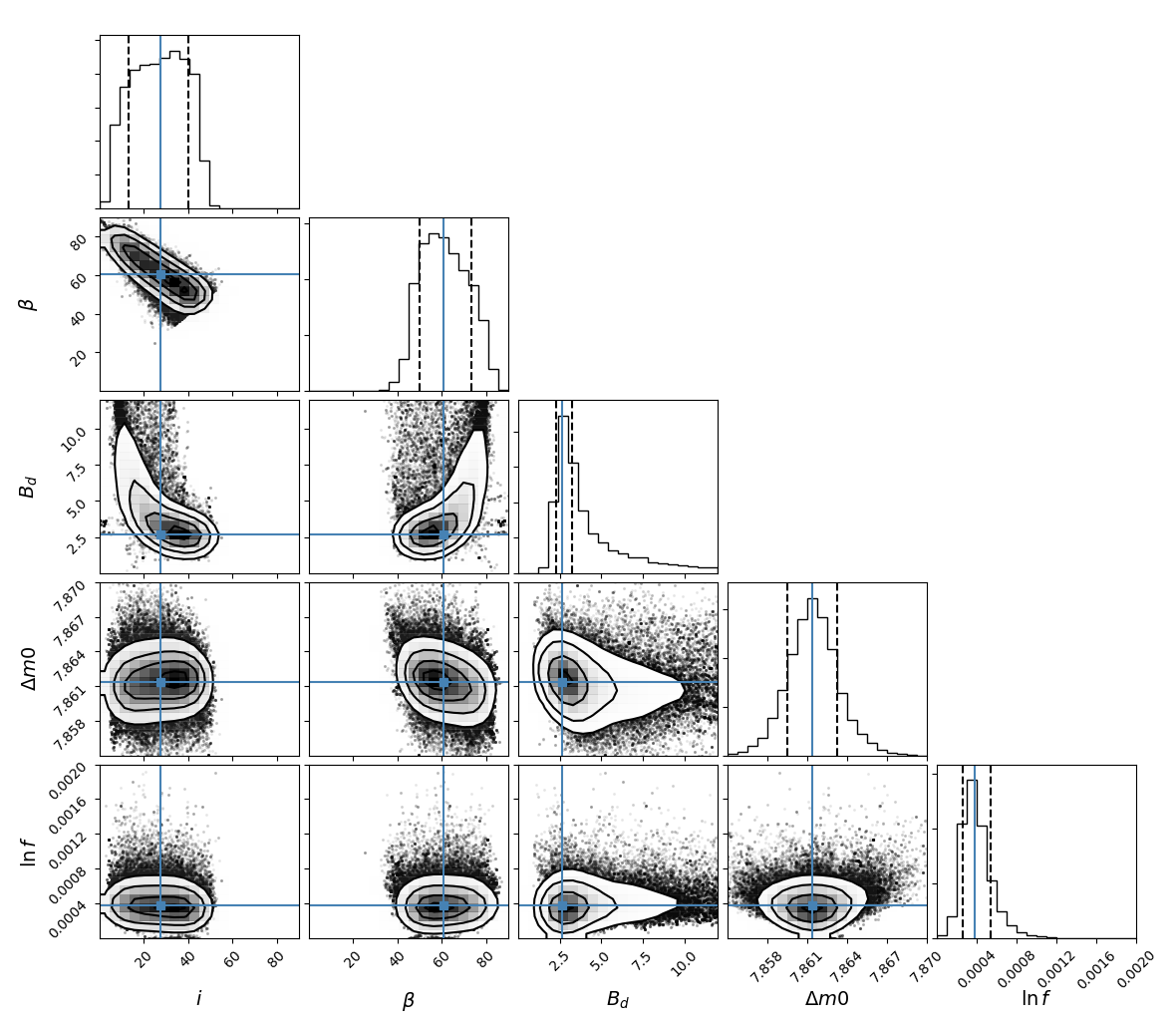}
	\caption{Likelihood distributions for the model parameters of HD 191612. Contours are drawn at the 16\%, 50\% and 84\% probability levels.}
	\label{fig:dist}
\end{figure}

\begin{figure}
	\includegraphics[width=\columnwidth]{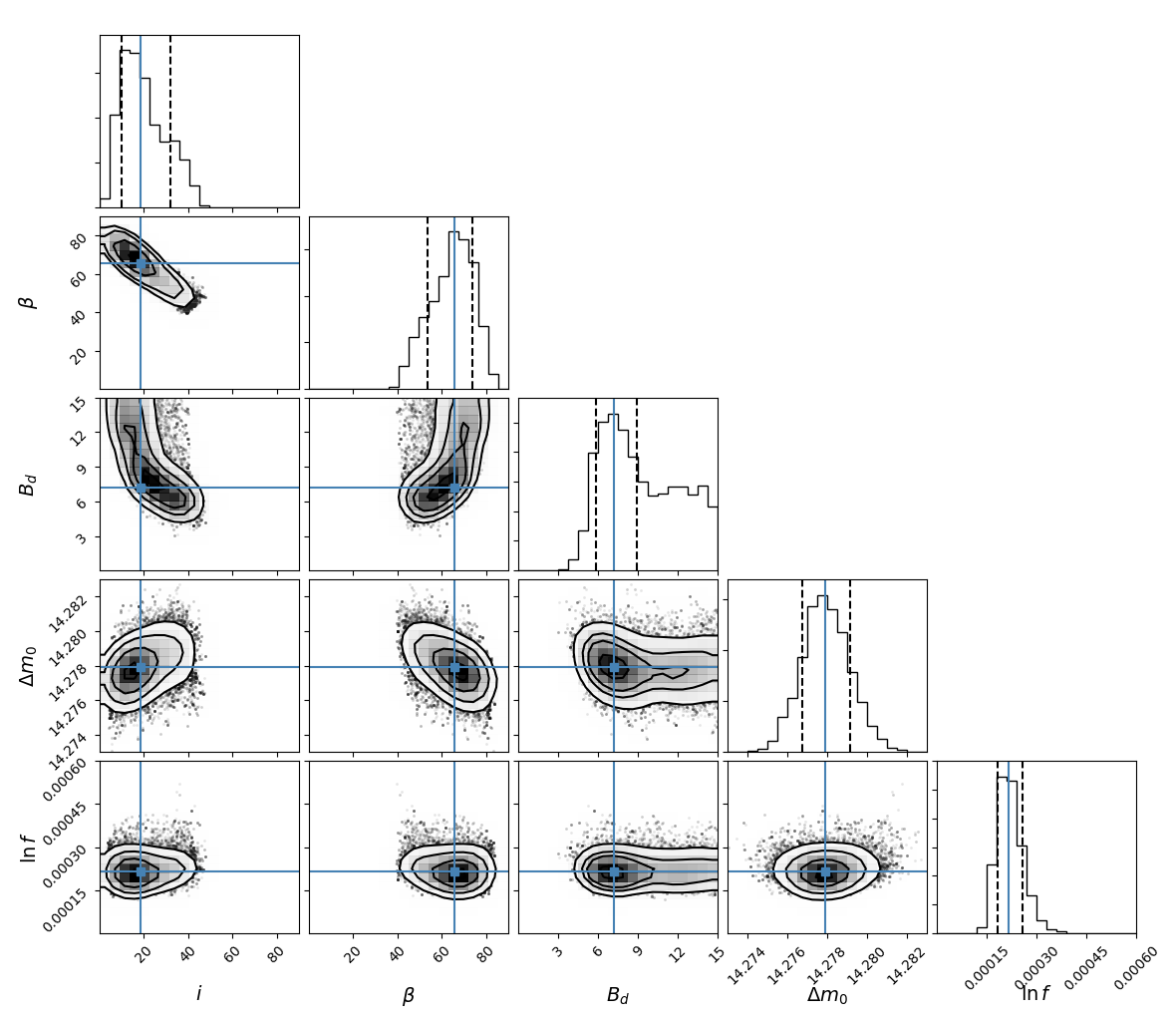}
	\caption{Likelihood distributions for the model parameters of 2dFS 936. Contours are drawn at the 16\%, 50\% and 84\% probability levels.}
	\label{fig:dist1}
\end{figure}

\begin{figure}
	\includegraphics[width=\columnwidth]{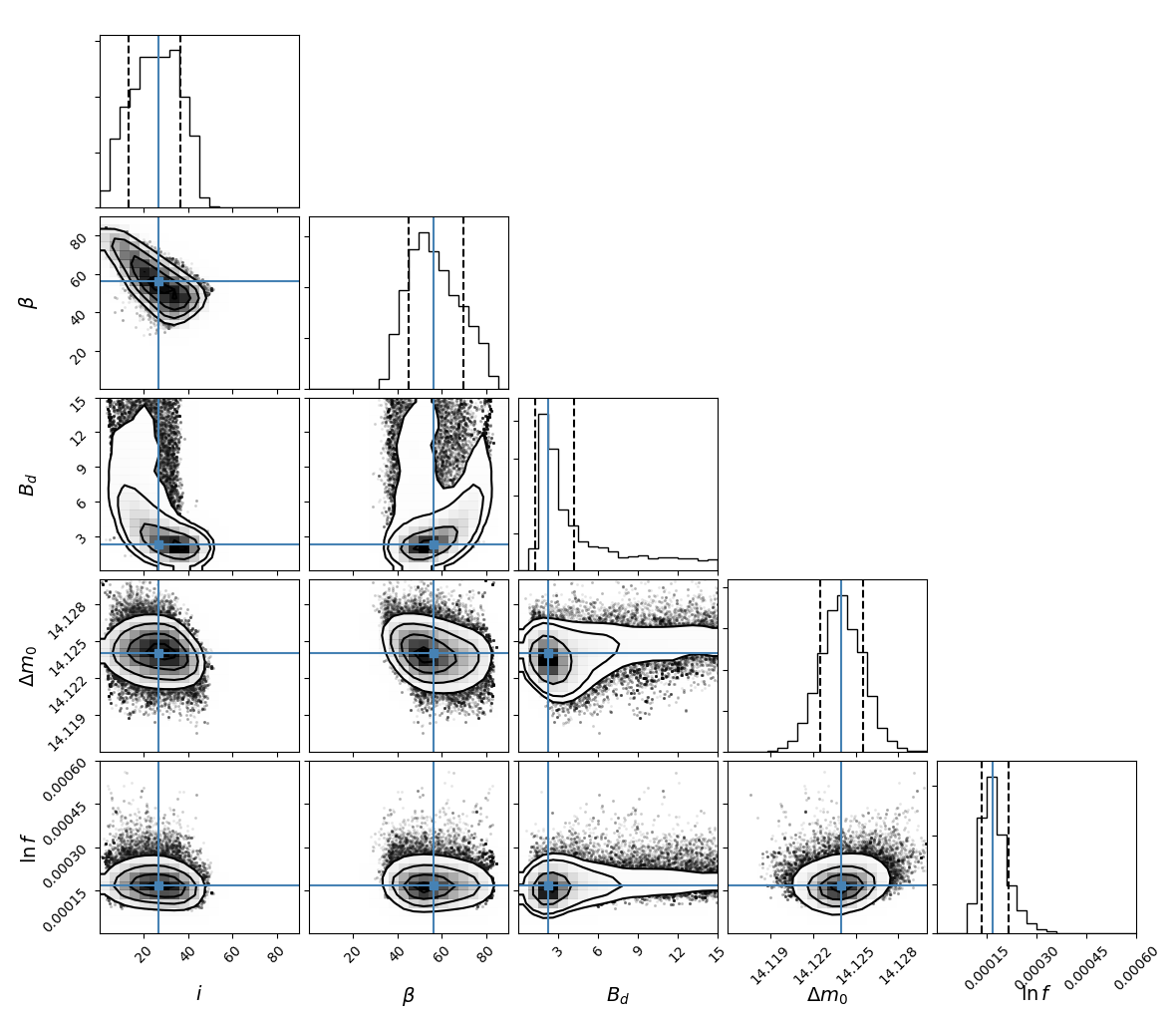}
	\caption{Likelihood distributions for the model parameters of BI 57. Contours are drawn at the 16\%, 50\% and 84\% probability levels.}
	\label{fig:dist2}
\end{figure}

\begin{figure}
	\includegraphics[width=\columnwidth]{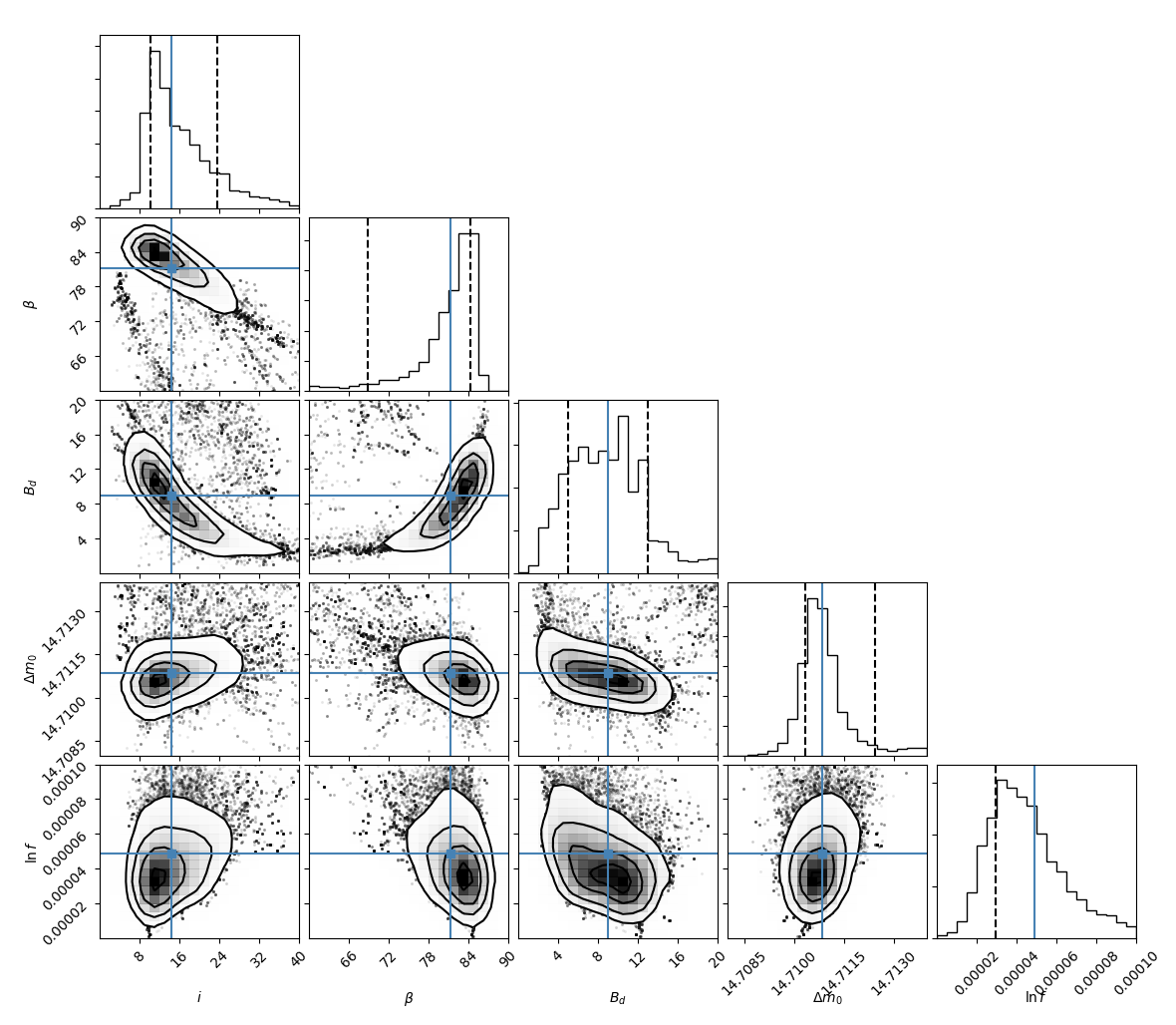}
	\caption{Likelihood distributions for the model parameters LMC164-2. Contours are drawn at the 16\%, 50\% and 84\% probability levels.}
	\label{fig:dist3}
\end{figure}

\begin{figure}
	\includegraphics[width=\columnwidth]{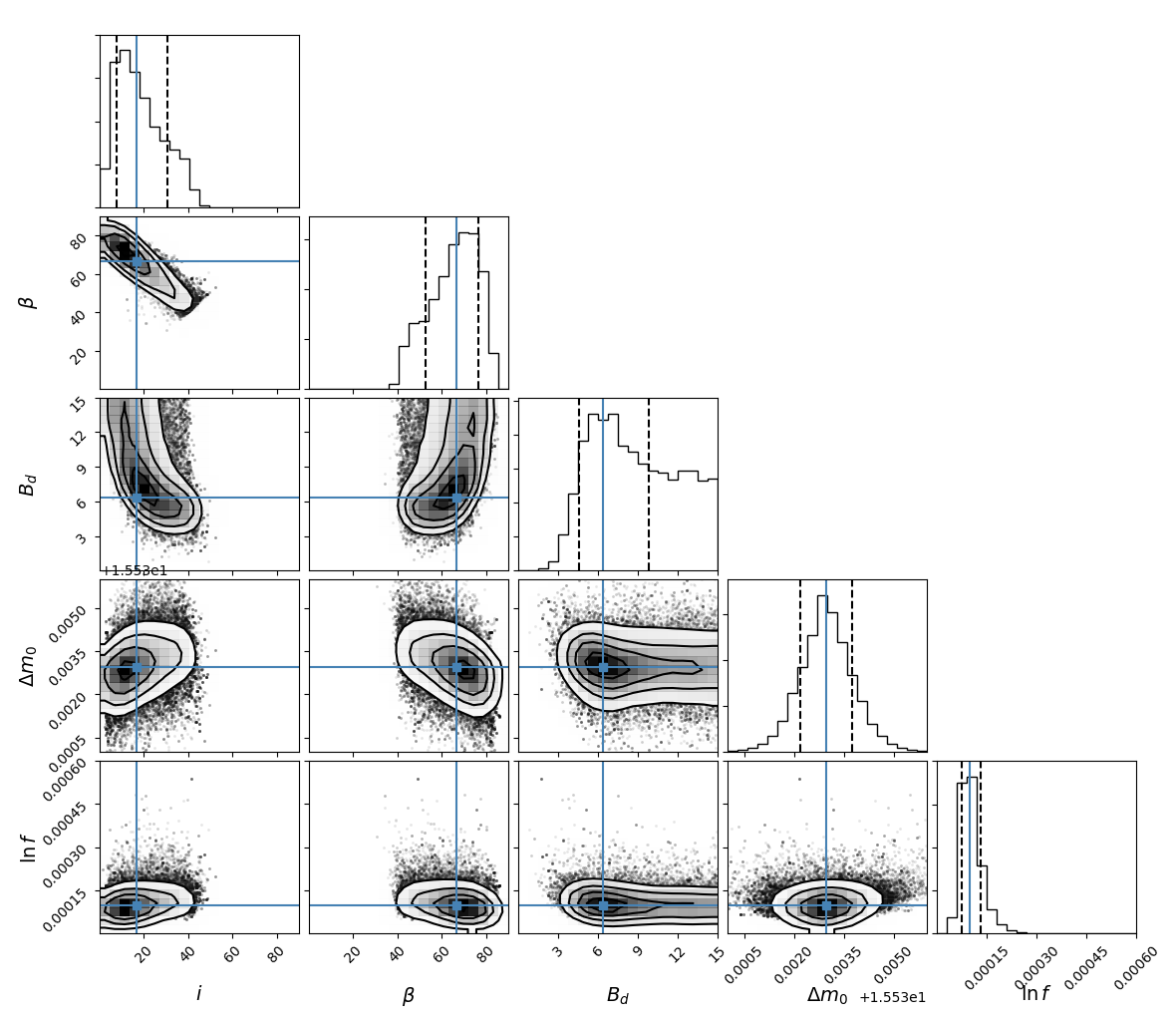}
	\caption{Likelihood distributions for the model parameters of SMC159-2. Contours are drawn at the 16\%, 50\% and 84\% probability levels.}
	\label{fig:dist4}
\end{figure}

\begin{figure}
	\includegraphics[width=\columnwidth]{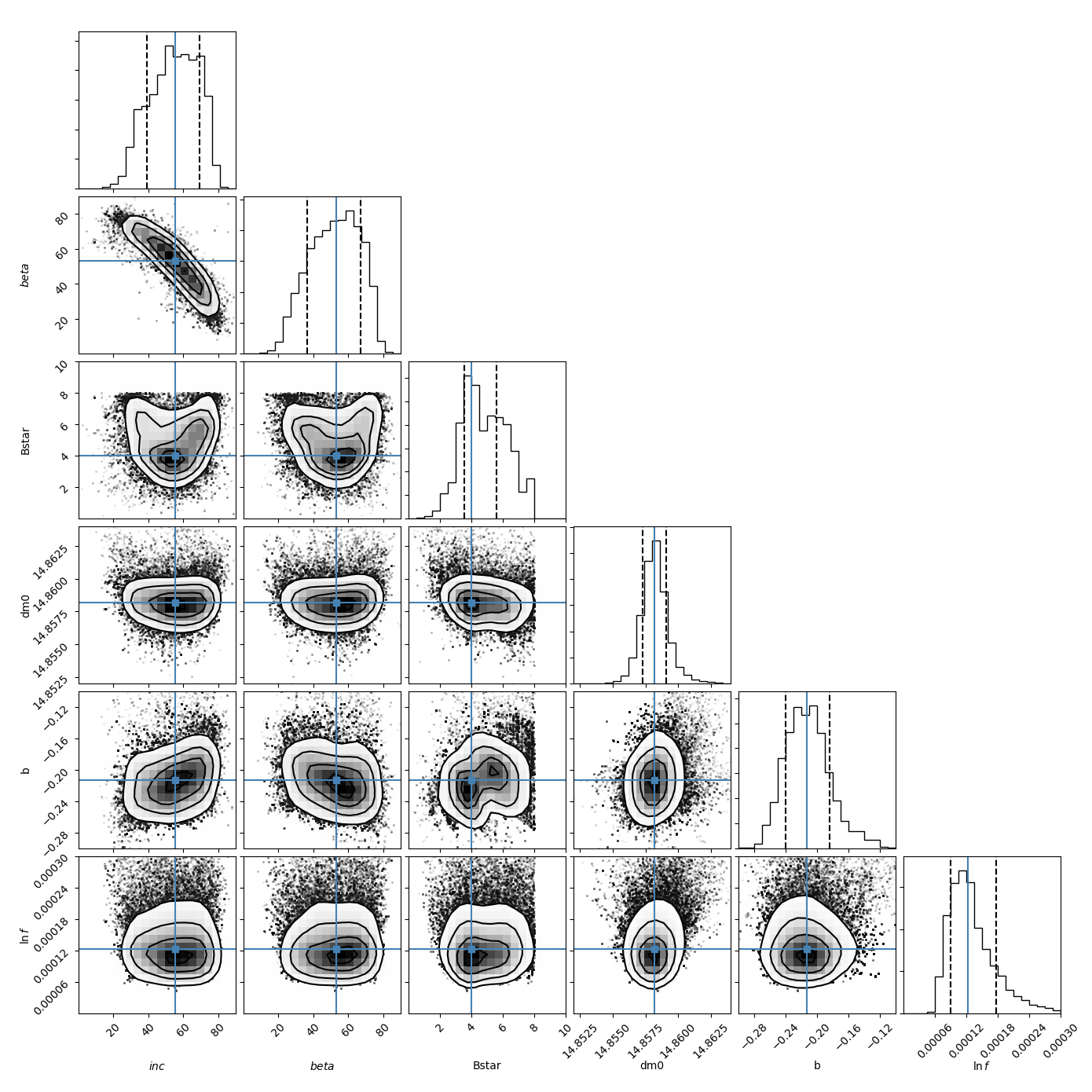}
	\caption{Likelihood distributions for the model parameters of LMCe136-1. Contours are drawn at the 16\%, 50\% and 84\% probability levels.}
	\label{fig:dist5}
\end{figure}


	\bsp	
	\label{lastpage}
	\end{document}